\documentclass[submission, Phys]{SciPost}
\usepackage{physics}
\usepackage{graphicx}
\usepackage{float}

\begin{document}

\begin{center}{\Large \textbf{
  Quantum fidelity susceptibility in excited state quantum phase
  transitions: application to the bending spectra of nonrigid molecules.
}}\end{center}

% TODO: write the author list here. Use initials + surname format.
% Separate subsequent authors by a comma, omit comma at the end of the list.
% Mark the corresponding author with a superscript *.
\begin{center}
J. Khalouf-Rivera\textsuperscript{1},
M. Carvajal\textsuperscript{1,2},
F. Pérez-Bernal\textsuperscript{1,2*}
\end{center}

\begin{center}
{\bf 1} Depto. de Ciencias Integradas y Centro de Estudios
  Avanzados en Física, Matemáticas y Computación, Universidad de
  Huelva, Huelva 21071, SPAIN
\\
{\bf 2} Instituto Carlos I de Física Teórica y Computacional,
  Universidad de Granada, Granada 18071, SPAIN
\\
* curropb@uhu.es
\end{center}

\begin{center}
\today
\end{center}

% For convenience during refereeing: line numbers
%\linenumbers

\section*{Abstract}
{\bf
We characterize excited state quantum phase transitions in the two dimensional
limit of the vibron model with the quantum fidelity susceptibility, comparing
the obtained results with the information provided by the participation
ratio. As an application, we perform fits using a four-body algebraic
Hamiltonian to bending vibrational data for several molecular species and, using
the optimized eigenvalues and eigenstates, we locate the eigenstate closest to
the barrier to linearity and determine the linear or bent character of the
different overtones.}

\vspace{10pt}
\noindent\rule{\textwidth}{1pt}
\tableofcontents\thispagestyle{fancy}
\noindent\rule{\textwidth}{1pt}
\vspace{10pt}

\section{Introduction}
\label{intro}
% Introduction:
% % QPT and ESQPT [V]
% % Hamiltonian with two symmetries [V]
% % Unitary groups U(n) [V]

The study of bending vibrational degrees of freedom has been fostered
due to their two-dimensional nature and the existence of two
well-defined physical limits --linear and bent configurations--,
together with intermediate configurations --quasilinear species--,
characterized by a large amplitude motion that makes them rich in
spectroscopic signatures~\cite{Quapp1993}. Positive or non-monotonous
anaharmonicities, the latter associated with the occurrence of the
Dixon dip in the Birge-Sponer plot for nonrigid molecules
\cite{Dixon1964}, and anomalous rotational spectra due to the mixing
of linear and bent characters in the wave functions of states
straddling in the propinquity of the barrier to linearity
\cite{Thorson1960,Kroto1992} are the most salient spectroscopic
features that can be found in the spectra of quasilinear species.

Significant advances and developments in spectroscopic methods have
made possible the experimental access to high bending overtones for
several molecular species. In this way, it has been possible to have
access to experimental spectroscopic information that allows for the
study of systems at energies around the barrier to linearity
\cite{Winnewisser2006, Reilly2015}. The results obtained for water
\cite{Zobov2005} and NCNCS \cite{Winnewisser2005, Winnewisser2010,
  Winnewisser2014} are of particular relevance.

In recent times, the concept of quantum monodromy, initially
introduced by Child \cite{Child1998}, has greatly helped in the assignment
of states in systems where the complexity of wave functions, due to
the proximity of the states to the barrier to linearity, hampered a
correct state labeling
\cite{Child1999,Winnewisser2005,Zobov2005,Winnewisser2006,Reilly2015}.
This is a concept borrowed from
classical mechanics that relies on the topological singularity
happening once the system energy is large enough to probe local saddle
points or maxima that prevent the definition of global action angle
variables \cite{Duistermaat1980}.

The theoretical modeling of bending vibrations in nonrigid molecular
species requires special tools, as the large amplitude vibrational
degree of freedom strongly couples vibrational and rotational degrees
of freedom.  A pioneering work in this field is the
Hougen-Bunker-Johns bender Hamiltonian \cite{HBJ}. This work was later
extended to the semirigid bender Hamiltonian \cite{Bunker1977} and the
general semirigid bender Hamiltonian \cite{Ross1988}. The MORBID model
\cite{MORBID}, based on the above mentioned developments, is currently a
standard method for the analysis of nonrigid molecular spectra, where
the simultaneous consideration of rotational and vibrational degrees
of freedom is required for the modeling of experimental term values
and the assignment of quantum labels.

The algebraic approach and, in particular, the vibron model is an
alternative to the traditional integro-differential approach for the
modeling of molecular spectra. This model is based upon symmetry
considerations and relies heavily in the properties of Lie algebras
\cite{Iachello2014}. The vibron model (VM) belongs to a family of
models that assign a $U(n+1)$ algebra as a dynamical or spectrum
generating algebra for an $n$-dimensional problem
\cite{general0}. Similar models have been successfully applied to the
modeling of the structure of hadrons \cite{Bijker1994,Bijker2000} and
nuclei \cite{booknuc,Iachello1991book,frank}. In the original vibron
model formalism, introduced by Iachello, rovibrational excitations of
diatomic molecular species are treated as collective bosonic
excitations \cite{Iachello:81}, and the dynamical algebra is $U(3+1)
=U(4)$, due to the vector nature of the relevant degrees of freedom
\cite{frank,bookmol}. The two-dimensional nature of bending vibrations
and the need to simplify the vibron model formalism to efficiently
deal with polyatomic systems, naturally drove to the formulation of
the two-dimensional limit of the vibron model (2DVM)
\cite{Iachello1996, PBernal2008}. The 2DVM defines a formalism that is
able to model the linear and bent limiting cases of the bending degree
of freedom, as well as the large amplitude modes that characterize
intermediate situations \cite{Iachello2003,PBernal2005, Larese2011,
  Larese2013}. An extension to four-body operators of the algebraic
Hamiltonian, used in the present work, has been recently published
\cite{KRivera2020}. The 2DVM has also been used for the modeling of
coupled benders \cite{Iachello1996, Iachello2008, Iachello2009,
  Larese2014}, stretch-bend interations \cite{Ishikawa2002,
  SCastellanos2012, Lemus2014, Marisol2020}, and the transition state
in isomerization reactions \cite{KRivera2019}.

In recent years, considerable attention has been paid to the occurrence of
quantum phase transitions (QPTs) in many different physical systems
\cite{Carr2010, Sachdev2011}. Such transitions occur at zero temperature and are
due to quantum fluctuations, differing in this way from the usual thermal phase
transitions. These transitions are also known as ground state quantum phase
transitions due to the abrupt modification experienced by the system
ground state wave function once a given parameter in the Hamiltonian (control
parameter) goes through a critical
value. The work on ground state QPTs in algebraic models can be traced back to
the seminal articles by Gilmore \textit{et al.} \cite{Gilmore1978, Gilmore1979,
  Feng1981} where such transitions were studied for nuclei. These transitions were also
called shape phase transitions as each phase corresponds with different
geometric configurations of the system's ground state. The study of QPTs in mesoscopic systems is a very active research line
\cite{Iachello2004,Casten2009, Cejnar2009, Cejnar2010} and a general
classification of QPTs in algebraic models can be found in
Ref.~\cite{Cejnar2007}. In the 2DVM case, the ground state QPT takes place
between the linear and bent limiting cases, with a second order phase transition
that occurs for nonrigid configurations \cite{Iachello2003,PBernal2005}. A full
ground state QPT analysis was performed in \cite{PBernal2008} and a study of
corrections beyond the mean field approach was published in \cite{PFernandez2011}. As this is
the simplest two-level algebraic model with a nontrivial angular momentum, it
has been chosen in many cases as a test model for QPT studies \cite{Zhang2010,
  Calixto2012, Calixto2012b, Santos2013, Castanos2015}.

The study of QPTs was later extended beyond the ground state with the concept of
excited state quantum phase transitions (ESQPTs) \cite{Cejnar2006, Cejnar2007b,
  Caprio2008}. Such transitions, often associated with a ground state QPT,
involve the non-analiticity of the energy level density and level flow for
critical values of the energy \cite{Fernandez2009, Fernandez2011b}. For a system
with \(n\) effective degrees of freedom, the order of the derivative of the level
density that is non-analytic is \(n-1\) \cite{Cejnar2008,
  Stransky2014,Stransky2015, Macek2019, Cejnar2020}.  In most cases, ESQPTs can be
associated with the existence of an unstable stationary point or a similar singularity in the potential
obtained in the classical limit of the system. The non-analiticity  is fully realized only in the system large size limit.
However,  ESQPT precursors can be easily identified for finite systems. Hence, in
an ESQPT there
exists a borderline of critical energy values, that marks the occurrence of a high
level density of states in a certain range of the control parameter or parameters. This line,  called separatrix, separates the different ESQPT phases. States
belonging to one of the phases have properties akin to the states of the
dynamical symmetry associated to the phase in question. As we discuss in this
work, it is often  cumbersome to assign a given excited state to a phase or
to ascertain its position relative to the separatrix. This is particularly complex for systems
with several control parameters and a complex phase diagram. ESQPTs have been
studied in different quantum many-body systems: the single \cite{Fernandez2009}
and coupled \cite{GRamos2017} Lipkin-Meshkov-Glick models, the Gaudin model
\cite{Relano2016}, the Tavis-Cummings and Dicke models \cite{Fernandez2011b,
  Brandes2013}, the interacting boson model \cite{Cejnar2009}, the kicked-top
model \cite{Bastidas2014}, periodic lattice models \cite{Dietz2013, Dietz2017},
or spinor Bose-Einstein condensates \cite{Feldmann2020,Cabedo2021}.  It has been paid
special heed to the influence of ESQPTs on the dynamics of quantum systems
\cite{Puebla2013,Puebla2015,Bastidas2015,Santos2015,Santos2016,PBernal2017,Wang2017,
  Kopylov2017, Kloc2018, Wang2019a, Wang2019b} and to link  ESQPT and
thermodynamic transitions \cite{BMagnani2016, PFernandez2017}. For a recent
review on the ESQPT subject, with a complete reference list, see
Ref.~\cite{Cejnar2020}.

The 2DVM presents an ESQPT, associated with a second order ground
state QPT, that can be explained from the influence on excited states
that have enough excitation energy to straddle the barrier to
linearity. There is a clear connection between the ESQPT phenomenon
and quantum monodromy \cite{PBernal2008}, something that can be
generalized to systems other than molecules \cite{Heinze2006,
  Macek2006, Kloc2017}. In fact, the possibility of accessing
highly-excited bending levels experimentally makes molecular
spectroscopy an optimal playground to detect ESQPTs precursors in
experimental spectra \cite{Larese2011, Larese2014, KRivera2019,
  KRivera2020}. Other systems where ESQPT signatures have been
experimentally recorded are superconducting microwave billiards
\cite{Dietz2013} and spinor Bose-Einstein condensates \cite{Zhao2014}.

As mentioned above, states lying at different sides of the separatrix
can be ascribed to one or the other of the existing limiting physical
situations, or dynamical symmetries. In the 2DVM case, as explained in
Sect.~\ref{sec2}, states can have a \(U(2)\) --linear-- or \(SO(3)\)
--bent-- character. However, as one gets further from the limiting cases
and closer to the critical energy, it gets cumbersome to assign states
to a given phase, due to the strong mixing in the wave function
\cite{KRivera2020}. This  explains by the known fact that the
definition of order parameters for ESQPTs is not an easy task, in
contrast with the situation for ground state QPTs
\cite{Caprio2008}.

Recently, a quantity called participation ratio \cite{Evers2008} (also
known as inverse participation ratio \cite{Izrailev1990} or number of
principal components \cite{Zelevinsky1996}), akin to the Shannon
entropy, has been used to quantify the degree of localization of
states when expressed in the bases for the different dynamical
symmetries. For systems with \(U(n+1)\) dynamical algebra and a second
order ground state QPT of the type \(U(n)-SO(n+1)\), it has been shown
that the participation ratio allows to reveal the location of the
ESQPT critical energy due to the enhanced localization of eigenstates
with energies close to the critical energy value if they are expressed
in the \(U(n)\) basis, \cite{Santos2015, Santos2016,
  PBernal2017}. This fact has been later confirmed, using the 2DVM, in
the study of the bending vibrational spectrum of molecular species
with large amplitude bending degrees of freedom \cite{KRivera2020} and
in the HCN-HNC isomerization transition state \cite{KRivera2019}. As
explained in Sect.~\ref{sec2}, the participation ratio does not allow
in all cases for an unambiguous assignment of a linear or bent
character to a given state. The large mixing that occurs once the
system is far enough from the dynamical symmetry limits hinders this
assignment, a fact that can be explained using the quasidynamical
symmetry concept \cite{Rowe2004}.

Therefore, it is important to find a quantity other than the participation ratio
that allows for the unambiguous assignment of 2DVM excited states around an ESQPT
to one of the implied phases. In recent times, quantum-information-derived quantities have been successfully employed to characterize ground state QPTs as they offer an approach that does not rely on the identification of an order parameter and its corresponding symmetry-breaking pattern (see \cite{Amico2008,Gu2010,Braun2018} and references therein). Inspired by these works, we have found that we can unambiguously assign excited states into ESQPT phases using quantum fidelity or quantum fidelity susceptibility.  Quantum fidelity is
a concept that arises in quantum information theory and it involves the overlap
of wave functions \cite{Nielsen2000}. This quantity has been successfully
applied to the study of ground state quantum phase transitions and critical
phenomena \cite{Zanardi2006}; for a review see \cite{Gu2010}.  A derived quantity that has been used for the
characterization of QPTs is the quantum fidelity susceptibility (QFS), the second derivative
of the fidelity and the leading order term in the series expansion of the
fidelity \cite{You2007,Gu2008,Gu2010}.  In the present work we extend the
calculation of quantum fidelity and QFS to 2DVM
excited states. We obtain an unambiguous assignment of such states to one
of the possible ESQPT phases, as we can locate the state position relative to the separatrix
between ESQPT phases. We apply the formalism to a recently presented
four-body 2DVM Hamiltonian  \cite{KRivera2020}, and we assign a
linear or bent character to the  excited states  of
linear and non-rigid molecular bending vibrations.

The present work is structured as follows. We provide a brief
introduction to the 2DVM in Sect.~\ref{sec2}, presenting a simple
model Hamiltonian and the two possible dynamical symmetries. In
Sect.~\ref{sec3}, we discuss the participation ratio and the quantum
fidelity results for the model Hamiltonian. In Sect.~\ref{sec4}, we
introduce the four-body Hamiltonian used to fit to experimental
data and we apply the formalism to the Si$_2$C molecule, a
well-known example of nonrigid molecule \cite{Reilly2015}. Finally, in an abridged form, we
also show results for selected bending degrees of freedom for other
five molecular species. In Sect.~\ref{sec5}, we include a summary of
the work and some concluding remarks.

%%%%%

\section{The two-dimensional limit of the vibron model}
\label{sec2}

Due to the two-dimensional nature of
bending vibrations, its algebraic modeling  implies the
treatment of vibrational quanta as collective bosonic excitations using the
$U(3)$ Lie algebra as a dynamical algebra  \cite{Iachello1996, PBernal2008}.

Following the algebraic formalism \cite{bookmol,frank}, one should
consider the possible dynamical symmetries that are subalgebra chains
starting in the dynamical algebra and ending in the system's symmetry
algebra. The conservation of angular momentum in a bending mode
(vibrational angular momentum) implies that the symmetry algebra in
this case is $SO(2)$. There exists two possible chains that start in
$U(3)$ and end in $SO(2)$
\footnote{A third chain, $U(3)\supset\overline{SO}(3)\supset SO(2)$ can be
  defined but it has the same physical interpretation than the
  $U(3)\supset SO(3)\supset SO(2)$ and it does not add new features to the model
  \cite{PBernal2008}.}
\begin{equation}\label{U3chains}
  \begin{array}{cccccl}
    U(3) &  \supset & U(2)    &  \supset         & SO(2)& \text{Chain (I)} \\
    N  &          & n       &                  & \ell & \\
    U(3) &  \supset & SO(3)   &  \supset         & SO(2)&\text{Chain (II)} \\
     N  &          & \omega  &                  & \ell & 
  \end{array}
\end{equation}
Each dynamical symmetry provides a set of quantum labels and a basis
to treat the problem of bending vibrational spectra and it is
associated with a limiting physical case. Chain (I) is known as the
cylindrical oscillator chain and it can be mapped with the bending
vibrations of a linear molecule. Its associated basis is a truncated
2D harmonic oscillator basis, with quantum labels \(\{|[N]\,
n^\ell\rangle\}\). The quantum label \(N\) identifies the totally
symmetric representation of \(U(3)\) and determines the size of the
system Hilbert's space; the \(n\) and \(\ell\) labels are the number
of quanta of excitation in the 2D harmonic oscillator and the
vibrational angular momentum, respectively. Chain (II) is known as the
displaced oscillator chain and it can be mapped to the limiting
physical case of a bent molecule. The associated basis is expressed as
\(\{|[N]\, \omega\, \ell\rangle\}\) where \(N\) has the same
interpretation than in the previous case, \(\omega\) can be connected
with \(\nu_b\), the number of quanta of excitation in the anharmonic
displaced oscillator: \(\nu_b = (N-\omega)/2\). Finally, \(\ell\), is
the projection of the molecular angular momentum on the figure axis.
In some cases this quantity is expressed using the usual notation for
symmetric tops \(\ell = K\).

The branching rules define the allowed
range of values for the quantum numbers in Eq.~(\ref{U3chains}).
\begin{equation}\label{BRules}
  \begin{array}{cccl}
    n = 0, 1, 2,\ldots N    &     & \ell = \pm n, \pm (n-2), \ldots , \pm 1 (\text{or } 0) &\text{Chain (I)}\\
    \omega = N, N-2,\ldots 1  (\text{or } 0)  &          & \ell = \pm \omega,
   \pm  (\omega-1), \ldots,\pm 1, 0 &\text{Chain (II)}\\
  \end{array}
\end{equation}
The calculations can be performed in any of the two basis, and the selection of
one or the other is often determined by the nature of the system under study. The
transformation brackets between both bases can be analytically derived
\cite{PBernal2008, Santopinto1996}.

In the algebraic approach, the Hamiltonian and any other operator of
interest is expressed as a function of Casimir or invariant operators
of the subalgebras in the different dynamical symmetries. A specially
convenient and simple model Hamiltonian can be built using two
operators: the number operator, \(\hat n\), and the pairing operator,
\(\hat P = N(N+1)-\hat W^2\). The first one is the first order Casimir
of the \(U(2)\) subalgebra while the second one is built with \(\hat
W^2\), the second order Casimir operator of the \(SO(3)\)
subalgebra. The model Hamiltonian,
\begin{equation}\label{modham}
  \hat{\cal H}(\xi) = \varepsilon\left[(1-\xi)\hat n + \frac{\xi}{N-1} \hat P\right]~,
\end{equation}
depends on two parameters, the system control parameter,
\(\xi\in[0,1]\), and the energy scale, \(\varepsilon\). Hereafter we
fix the energy scale to \(\varepsilon = 1\) and the calculated
energies are dimensionless quantities. The pairing operator is a
two-body operator, while the number operator is a one-body operator;
therefore the two-body part is normalized by the system size to make
the Hamiltonian intensive and allow for the calculation of results in
the thermodynamic or large size --large \(N\)-- limit. The study of
the eigenvalues of Hamiltonian (\ref{modham}) and its classical limit
determines that there exists a ground state QPT of second order with a
critical control parameter value \(\xi_c = 0.2\) \cite{PBernal2008}.
For control parameter values \(\xi\le\xi_c\), the system is said to be
in the \(U(2)\) or symmetrical phase, which in the molecular case can
be mapped to a linear configuration. If \(\xi>\xi_c\), then the system
is in a \(SO(3)\) or broken symmetry phase, known as a bent --or
semirigid-- configuration in the case of vibrational bending.

The conservation of
vibrational angular momentum implies that Hamiltonian (\ref{modham}) is block
diagonal in the quantum label \(\ell\). The nonzero matrix elements  in the
cylindrical oscillator basis are
%\begin{widetext}
\begin{align}
\langle [N] n_2^\ell |\hat {\cal H}(\xi)| [N] n_1^\ell\rangle =& \left[(1-\xi)
  n_1 + \frac{\xi\left\{N(N+1) - (N-n_1)(n_1+2) - (N-n_1+1)n_1 - \ell^2\right\}}{N-1}\right] \delta_{n_2,n_1}\nonumber\\
&+\frac{\xi}{N-1} \sqrt{(N-n_1+2)(N-n_1+1)(n_1+\ell)(n_1-\ell)}\,\delta_{n_2,n_1-2}  \\
&+\frac{\xi}{N-1} \sqrt{(N-n_1)(N-n_1-1)(n_1+\ell+2)(n_1-\ell+2)}\,\delta_{n_2,n_1+2}~~.\nonumber
\end{align}
%\end{widetext}

\begin{figure}
    \centering
    \includegraphics[%trim = 0 240 0 280, clip,
    width=1.0\textwidth]{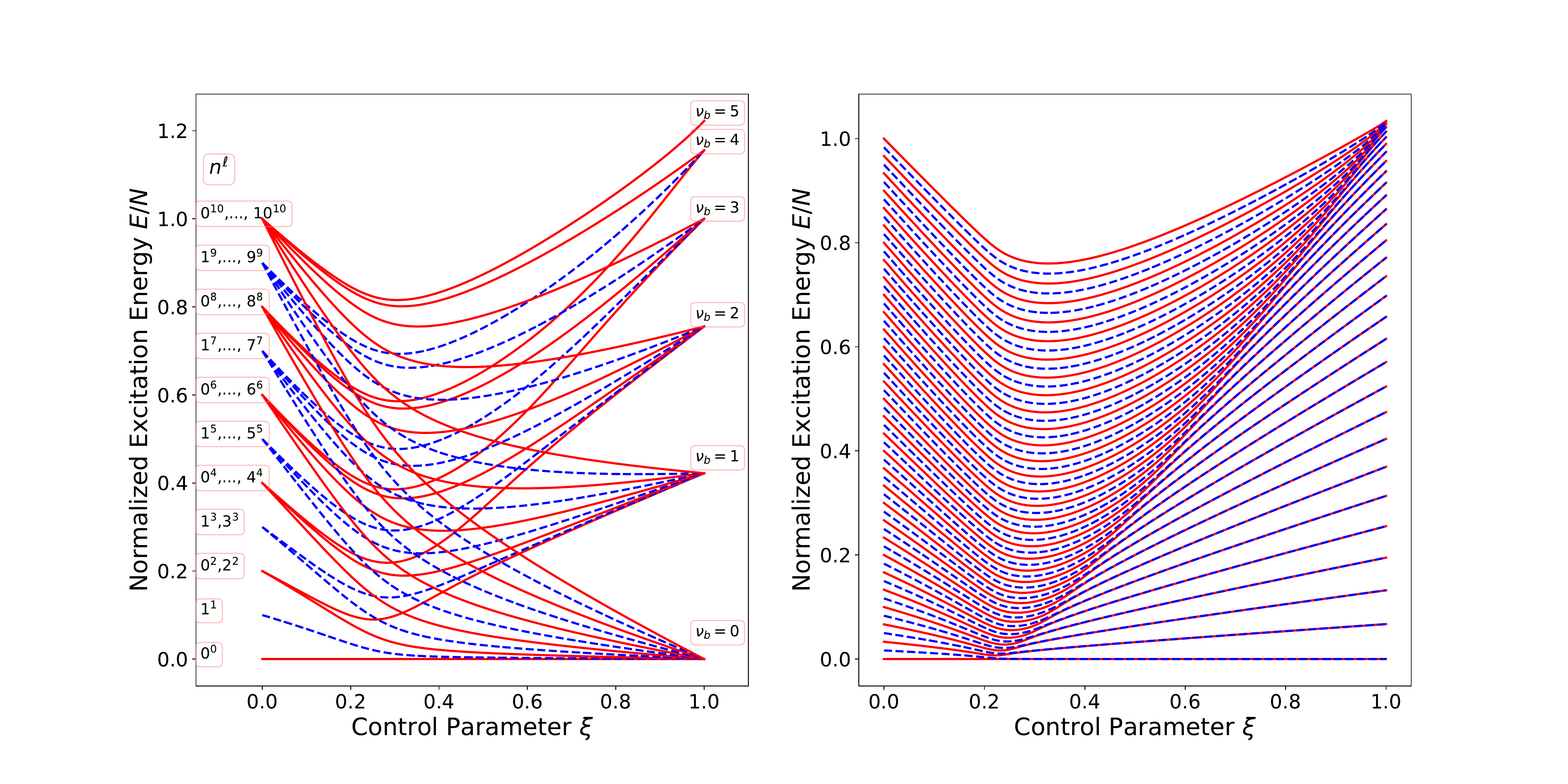}
    \caption{\label{figmodH} Normalized excitation energy of
      the 2DVM Hamiltonian (\ref{modham}) as a function of the control
      parameter $\xi$.  Left panel: system size is $N=10$ and all angular momenta states
      are shown (\(\ell = 0\ldots 10\)) using full red (dashed blue) lines for
      even (odd)
      angular momentum values. Right panel: system size is $N = 80$ and states with
      \(\ell = 0\) (\(\ell = 1\)) states are depicted using full red (dashed blue) lines. }
  \end{figure}

If \(\xi = 0\), the Hamiltonian is diagonal in Chain (I) basis and the
spectrum is harmonic; while in the \(\xi = 1\) case the Hamiltonian is
diagonal in Chain (II) basis and the spectrum is anharmonic, with
degenerate rotational bands. This can be clearly seen in the
correlation energy diagram depicted in the left panel of
Fig.~\ref{figmodH}, where the excitation energy is plot as a function
of the control parameter \(\xi\) for \(N =10\) using full red lines
for levels with even angular momentum values and dashed blue lines for
odd angular momentum levels. This plot allows to track level paths
from one dynamical symmetry to the other; from a 2D harmonic spectrum
on the left side (\(\xi = 0\)) to the anharmonic oscillator spectrum
in the right side (\(\xi = 1\)). We have included the quantum labels
\(n^\ell\) in the \(U(2)\) limiting case and the number of quanta of
excitation \(\nu_b\) associated to the \(\omega\) quantum label in the
\(SO(3)\) case. In the latter case, once \(\xi = 1\), values with
different vibrational angular momentum form a degenerate rotational
band. In between these two cases, the model Hamiltonian
(\ref{modham}), despite its simplicity, is able to reproduce spectra
with positive anharmonicity, associated with flat potentials for
control parameter values less than \(\xi_c\). It can also reproduce
the spectroscopic signatures of nonrigid molecular species: the Dixon
dip and the change from a linear to a quadratic dependence of the
energy with vibrational angular momentum that characterizes quantum
monodromy \cite{Iachello2003, PBernal2008, Larese2011, Larese2013,
  KRivera2020}.

%\begin{widetext}
%\end{widetext}

We also plot the correlation energy diagram for Hamiltonian (\ref{modham}) in
the right panel of Fig.~\ref{figmodH}, this time for a larger system size (\(N =
80\)) and including only \(\ell = 0\) (full red lines) and \(\ell = 1\) (dashed
blue lines) states. In this case, it is evinced a line marked by a locally high
density of energy levels, starting at \(\xi = \xi_c = 0.2\) and zero excitation energy. This line marks the ESQPT critical energy and
the boundary between the two ESQPT phases. It can also be easily appreciated
that in the \(SO(3)\) phase, below the line, states with different vibrational angular momenta are degenerate and
this degeneracy is broken in the \(U(2)\), for states above the line.

%%  in the intro note that the level density could be non analytic or a derivative of this quantity

\section{Quantum Fidelity Susceptibility in the 2DVM}\label{sec3}
%% Expression: when it comes to -ing -> a la hora de ...

A convenient tool for the characterization of wave functions in the
phases defined by an ESQPT is the participation ratio (PR). This quantity provides
 the degree of localization of a state in the available bases \cite{Evers2008} --it is also known as inverse
participation ratio \cite{Izrailev1990} or number of principal
components \cite{Zelevinsky1996}. For a quantum  state $\ket{\Psi}$, expressed in a
given basis $\left\{\ket{\phi_i}\right\}_{i=1,...,\text{dim}}$ 
as $\ket{\Psi}= \sum\limits_{i=1}^{\text{dim}} c_i \ket{\phi_i} $, the PR is defined as
\begin{equation}
  PR[\Psi]=\frac{1}{\sum\limits_{i=1}^{dim}
    \left|c_{i}\right|^4}~.
  \label{PRdef}
\end{equation}
Note that the minimum value of the PR for a given state is one, and this means
that the state under scrutiny belongs to the basis. On the other hand, the maximum
value is equal to the basis dimension, $dim$, in the --nonrealistic-- case of a state with equal
and non zero $c_i$ coefficients ($c_i = 1/\sqrt{dim}$).

In algebraic models with an ESQPT associated with \(U(n)-SO(n+1)\)
dynamical symmetries,i t has been found that state(s) close to the
critical energy of the ESQPT display a high localization in one of
dyanamical symmetry basis \cite{Santos2015, Santos2016, PBernal2017}.
The main application of this quantity in the 2DVM stems from the high
localization of the $\ell = 0$ states that lie closer to the barrier
to linearity --critical energy of the ESQPT-- when expressed in the
the $U(2)$ basis (\ref{U3chains}).  More precisely,
such state, or states, have a dominant component for the basis state
$|[N] n=0^{\ell =0}\rangle$ \cite{KRivera2020}. This effect is blurred for increasing
values of the vibrational angular momentum, $\ell$. This is an effect that can be explained by  the centrifugal
barrier precluding the wave function from exploring the barrier to
linearity critical point. The PR has also proved useful in the caracterization of different type
of ESQPT, associated with the $\ell = 0$ transition states in
isomerization reactions, where the $|[N] n=N^{\ell =0}\rangle$
component has the highest weight in the Chain (I) basis \cite{KRivera2019}.

In the 2DVM, cases with an ESQPT --i.e.~with  quantum monodromy--, notwithstanding the critical energy of the
ESQPT is well determined from the PR values for eigenstates in the \(U(2)\) basis,
the comparison of the PR values obtained for the \(U(2)\) and the  \(SO(3)\)  bases does not
allow for a clear assignment of a given eigenstate to a linear or bent ESQPT
phase. This is specially relevant for systems with a low barrier to linearity, and for states
that lie far from both
limiting physical cases, the \(U(2)\) and \(SO(3)\) dynamical symmetries. This 
is in good accordance with the quasidynamical symmetry concept \cite{Rowe2004}, that
explains the high degree of mixing expected as one gets further from the
dynamical symmetries, even for states retaining most of the characteristic features of
a dynamical symmetry. Therefore, in such cases, the direct comparison of the PR
values for the \(U(2)\) and \(SO(3)\) bases does not allow an unambiguous
assignment of the eigenstate to a linear or bent character. 

Thus, we have looked for a basis-independent quantity that could
achieve an unambiguous assignment of a system excited states to one of
the existing phases in a precise manner. This is specially relevant
once we move from the simple model Hamiltonian (\ref{modham}) to more complex Hamiltonians that 
include higher order operators.

Our proposal is to extend the QFS to 2DVM excited
eigenstates,  obtaining in this way a sensitive
probe, able to locate a given eigenstate position with respect to the separatrix line
between ESQPT phases. It  therefore allows to assign excited states to a
\(U(2)\) or \(SO(3)\) ESQPT phase in a basis-independent way. We proceed to define QFS and its
application to the 2DVM.

The definition of quantum fidelity, a quantity introduced in quantum information theory
\cite{Nielsen2000}, for a system with a single control parameter, $\lambda$, is
\begin{equation}
  F(\lambda,\delta\lambda) =
  \left|\bra{\psi_0(\lambda)}\ket{\psi_0(\lambda + \delta\lambda)}\right|~.
\end{equation}
This quantitiy provides a measure of the similarity between ground quantum
states obtained for control parameters values $\lambda$ and $\lambda +
\delta\lambda$. Despite its apparent simplicity, this quantity efficiently grasps the sudden change
experienced by the ground state wave function once the control parameter is
varied across its critical value and, since the seminal work of Zanardi \cite{Zanardi2006},
it has been used to characterize QPTs in different systems
\cite{You2007,Gu2010}. Another magnitude often used to identify QPTs is the
QFS \cite{You2007,Gu2010}, which is maximum when the parameter
\(\lambda\) goes through a critical value
\begin{equation}
  \chi_{F}(\lambda)= -\frac{\partial^2F(\lambda,\delta\lambda)}{\partial(\delta\lambda)^2} = \lim_{\delta\lambda\to 0}\frac{-2\ln{F(\lambda,\delta\lambda)}}{(\delta\lambda)^2}~.\label{chi_F}
\end{equation}
Using perturbation theory, the QFS can be expressed  in the so called summation
form \cite{Gu2010}
\begin{equation}
\chi_{F}(\lambda)= \sum^{dim}_{i\ne0}\frac{\left|\bra{\psi_i(\lambda)} \hat H^I\ket{\psi_0(\lambda)}\right|^2}{\left[E_i(\lambda)-E_0(\lambda)\right]^2}~~,\label{fidsuscep}
\end{equation}
\noindent where \(\hat H^I\) is the interaction Hamiltonian and the total
Hamiltonian can be written as \(\hat H(\lambda) = \hat H^0 + \lambda H^I\);
\(\ket{\psi_i(\lambda)}\) is the \(i\)-th eigenvector of Hamiltonian \(\hat
H(\lambda)\) and \(E_i(\lambda)\) is its eigenvalue. An important advantage
of the QFS, expressed in this form, is that it is
independent of the \(\delta\lambda\) value. 

The QFS has been used in the characterization of ground
state quantum phase transitions and their universality in relevant
many-body quantum systems, e.g.~the 1D Hubbard model \cite{You2007,
  Carrasquilla2013}, the Kitaev honeycomb model \cite{Yang2008}, the 1D
asymmetric Hubbard model \cite{Gu2008}, the Lipkin-Meshkov-Glick model
\cite{Kwok2008, Leung2012, Romera2014}, the two-dimensional transverse-field Ising and XXZ
models \cite{Yu2009}, the Rabi model \cite{Wei2018}, Gaussian random ensambles \cite{Sierant2019}, or 1D lattice models
\cite{Rams2011,Greschner2013, Wei2020, Leblond2020}.

In the present work, we  extend the concept of QFS beyond the ground state, to
the realm of excited states, and we use this
magnitude as a probe to locate excited states in the 2DVM with respect to the
separatrix line between different ESQPT phases. We will apply this to the
results obtained in the fit of Hamiltonian (\ref{H4b}) to several molecular
species obtaining an unambiguous assignment of the excited states to a given
basis.

Our proposal is to introduce a control parameter $\lambda$ and split the
algebraic 2DVM spectroscopic Hamiltonian into three different terms: a first one,
$\hat{H}_{I}$, that encompasses all operators  diagonal in the \(U(2)\) basis and its associated spectroscopic parameters; a
second one, $\hat{H}_{II}$, including terms diagonal in the \(SO(3)\) basis; and a
third one,  $\hat{H}_{I-II}$,
containing operators and the corresponding spectroscopic parameters diagonal in both bases
\begin{align}
  \hat{H} \left(\lambda\right) = & \left(1-\lambda\right) \hat{H}_{I} +
                                   \left(1+\lambda\right) \hat{H}_{II} +
                                   \hat{H}_{I
                                   -II}= \hat{H}(\lambda=0)  + \lambda \hat{H}^I~,\label{Hlamb}\\
  \hat{H}^I         = &                       -\hat{H}_{I} +\hat{H}_{II} ~.\label{HI}
\end{align}

The control parameter \(\lambda\) is defined in the range \(\lambda\in[-1,1]\) and the initial Hamiltonian is recovered for $\lambda =0$. The Hamiltonian \(
\hat{H} \left(\lambda=\pm 1\right)\) is diagonal in the \(U(2)/SO(3)\) basis.

We now proceed to define the QFS for the  \(j-\)th
eigenstate of Hamiltonian \(\hat{H} \left(\lambda\right)\) as 
\begin{equation}
\chi^{(j)}_{F}(\lambda)= \sum^{dim}_{i\ne j}\frac{\left|\bra{\psi_i(\lambda)} \hat H^I\ket{\psi_j(\lambda)}\right|^2}{\left[E_i(\lambda)-E_j(\lambda)\right]^2}~,\label{fidsuscepII}
\end{equation}
\noindent that is a generalization of Eq.~(\ref{fidsuscep}) to excited states. As the value of the \(\lambda\) control parameter is varied,
 \(\chi^{(j)}_{F}(\lambda)\) will evidence --even for
finite-size systems-- a peak whenever a separatrix line associated with an ESQPT
is crossed. A
similar procedure has been recently published, using QFS of excited states, in
the study of the adiabatic and counter-adiabatic driving in ESQPTs \cite{Cejnar2020} and of the onset of quantum chaos in spin chain models
\cite{Leblond2020}.

\begin{figure}
    \centering \includegraphics[trim = 0 220 0 280, clip,
    width=1.0\textwidth]{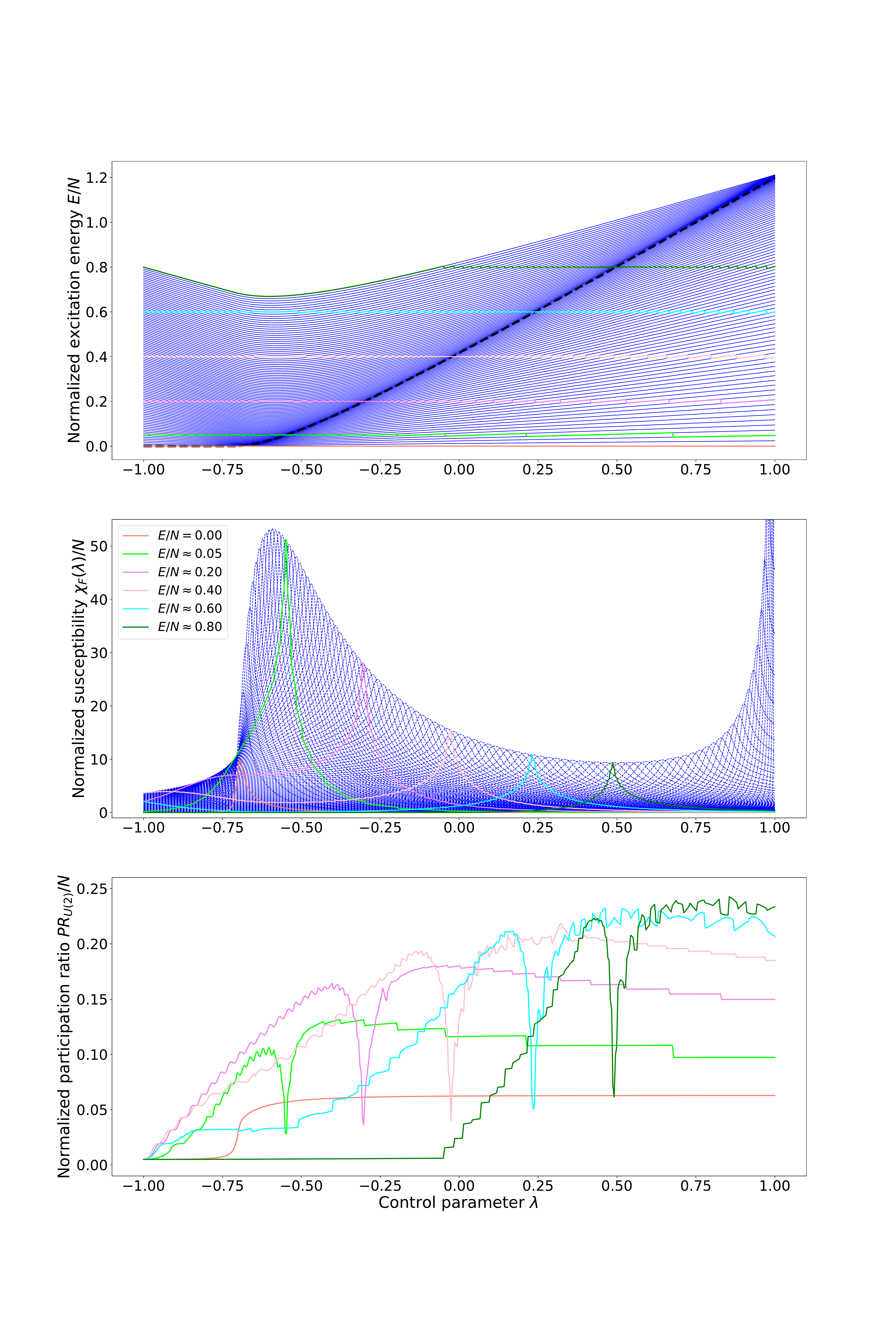}
    \caption{\label{mHsus} All panels: Results for a 2DVM Hamiltonian
      (\ref{modham1}) with $\xi=0.6$ and $N=200$. For the sake of
      clarity, the ground state and states with normalized excitation
      energies closer to $0.05, 0.2, 0.4, 0.6$ , and $0.8$ have been
      highlighted using different colors (orange, light green, purple,
      pink, cyan, and dark green, respectively). The rest are plotted
      with blue dashed lines. All quantities are plotted versus the
      control parameter $\lambda$. Upper panel: Normalized excitation
      energy of \(\ell = 0\) eigenstates of the model Hamiltonian
      (\ref{modham1}). Middle panel: Normalized QFS
      (\ref{fidsuscepII}). Lower panel: Normalized PR in the $U(2)$
      basis  (\ref{PRdef}) for the set of selected states indicated above.}
  \end{figure}

We show as an example the application of Eq.~(\ref{fidsuscepII}) to the excited
states of model Hamiltonian (\ref{modham}) for a fixed  \(\xi>0.2\). In this case, \(\hat{H}_{I}(\xi) =
(1-\xi)\hat n\) and \(\hat{H}_{II}(\xi) = \xi/(N-1)\hat P\), and
\begin{equation}
  \hat{H}\left(\lambda\right) = \hat{\cal H}(\xi) + \lambda \hat{H}^I(\xi)~ ,\label{modham1}
\end{equation}
\noindent where \(\hat{H}^I(\xi) = -\left(1-\xi\right)\hat{n} +
\left(\frac{\xi}{N-1}\right)\hat{P}\) and the new control parameter is
\(\lambda\).  We show the results obtained for \(\ell=0\) states of
the model Hamiltonian with a control parameter value $\xi=0.6$ and a
system size $N = 200$ in Fig.~\ref{mHsus}. The correlation energy
diagram, plotting the normalized excitation energy versus the
\(\lambda\) control parameter, is shown in the upper panel. The
resulting diagram is, as expected, similar to the correlation energy
of $\hat{\cal H}(\xi)$, with a ground state QPT and a line of high
density of states that marks the ESQPT separatrix. The energies for
\(\lambda = 0\) are the energies of our selected model Hamiltonian
case. We have highlighted the results obtained for the ground state
and the states with normalized excitation energies closer to $0.05,
0.2, 0.4, 0.6$ , and $0.8$, with different colors (orange, light
green, purple, pink, cyan, and dark green, respectively).  Therefore,
instead of going across the ESQPT following a given eigenstate, we
have selected a set of states according to their excitation energy
values. For each one of them, the ESQPT separatrix is crossed at
different \(\lambda\) values (see upper panel).

We plot in Fig.~\ref{mHsus} center panel the results obtained for the QFS
(\ref{fidsuscepII}) normalized by the system size for the model Hamiltonian excited
states as a function of  \(\lambda\). We use the same color
code than in the upper panel to emphasize the results for a selected set of
states. It is
clear that the QFS for an excited state reaches its
maximum value when the state energy straddles the ESQPT critical energy
line. Therefore, if the maximum of the QFS for a
level occurs for a negative (positive) \(\lambda\) value, the level lies below
(above) the ESQPT separatrix and we can assign a \(SO(3)\) (\(U(2)\)) character
to the excited state. In case the QFS maximum value is obtained for \(\lambda = 0\), the system excited state energy
coincides with the critical energy and the state is in the separatrix
line. In the provided example, the ground state and the states with normalized energy values close to  \(0.05, 0.2\), and \(0.4\) are of
bent (\(SO(3)\)) type and the states with \(E/N\) close to  \(0.6\) and \(0.8\) have a linear (\(U(2)\))
character. The  eigenstate \(\nu_b = 50\) is the state with an energy closest to the separatrix for
\(\lambda = 0\).

The lower panel of  Fig.~\ref{mHsus} shows, for the same selected states and with the same color code, the value of the normalized PR in the $U(2)$ basis as a function of
the \(\lambda\) control parameter. In the ground state case, there
is an abrupt change in the PR value for the  \(\lambda\) value associated with
the ground state quantum phase transition, while 
excited states show a minimum in the participation ratio for the \(\lambda\)
control parameter value that makes them cross the ESQPT
separatrix, as predicted in Refs.~\cite{Santos2015,Santos2016,PBernal2017}.

\begin{figure}
    \centering \includegraphics[trim = 0 120 0 160, clip,
    width=1.0\textwidth]{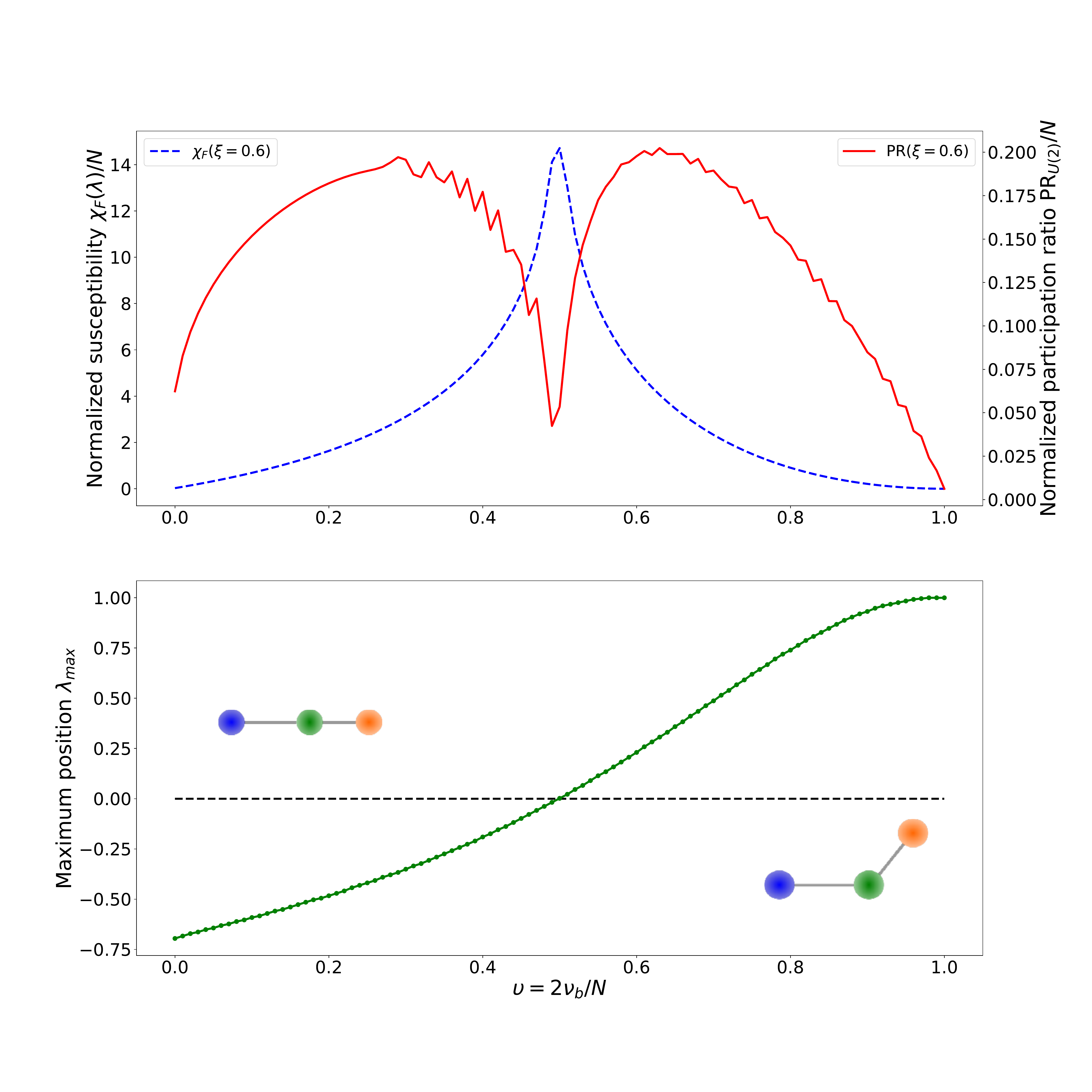}
    \caption{\label{mHlmax} All panels: Results for the 2DVM model Hamiltonian
      (\ref{modham1}) with  $\xi=0.6$ and $N=200$. Upper panel: PR (red solid
      line, right ordinate axis scale) and QFS (blue dashed line, left ordinate axis scale) for the $\lambda = 0$
      eigenstates. Lower panel: Position of the QFS
      maxima, $\lambda_{\text{max}}$, for each eigenstate versus
      the normalized bending quantum number $2\nu_b/N$. The
      black dashed line marks the $\lambda=0$ value. }
  \end{figure}

In order to further illustrate the role of the QFS,
we plot in the upper panel of Fig.~\ref{mHlmax} 
the normalized QFS --blue dashed line, left ordinate axis scale--
and PR --red solid line, right ordinate axis scale-- as a function of the
normalized excitation number (\(2\nu_b/N\)) for the model
Hamiltonian eigenstates, with \(\lambda = 0\), \(\xi = 0.6\), and \(N = 200\). Both
quantities can be used to assess the value of the ESQPT critical energy,
indicated by a minimum (maximum) value of the PR (QFS).

The lower panel of Fig.~\ref{mHlmax} shows $\lambda_{\text{max}}$, the value of
the control parameter $\lambda$ for which each eigenstate of the model
Hamiltonian with  \(\xi = 0.6\) and \(N = 200\) has a maximum QFS
value as a function of the normalized excitation number. The horizontal black
dashed line marks the $\lambda=0$ value and, as previously stated, eigenstates with
$\lambda_{\text{max}}<0$ ($\lambda_{\text{max}}>0$) can be classified as linear- (bent-like) states.

%\begin{widetext}
%\end{widetext}

%\begin{widetext}

  Therefore, the QFS provides a trustworthy and
  basis-independent method to locate states with respect to the high level
  density separatrix lines that characterize ESQPTs. The case (\ref{modham1}) is a
  particularly simple one, but in the next section we show how to use the
  QFS in a more general case, with an application to the
  bending wavefunctions obtained from the fit of a  Hamiltonian including up to four-body interactions
  to reported vibrational bending band origins for different
  molecular species. In fact, the difficulty of clearly assigning
  levels in cases such as the ones included in the next section has been the original motivation for this
  research \cite{KRivera2020}.

%\end{widetext}
\section{Application to molecular bending structure}
\label{sec4}
% Results
% % Extended Si2C section explaining all issues.
% % Global results and H(4b) fits
% % Global susceptibility graphs
% % Reduced quasilinear and nonrigid subsections of the rest of molecules
In order to illustrate how PR and QFS can help in
the characterization of bending vibrational excited states we
apply the procedure explained in section~\ref{sec3} to reported data for several
molecules. We have selected mostly nonrigid species, due to their feature-rich bending
spectrum that includes an ESQPT once eigenstates straddle the barrier to linearity.  In particular, we show results for
bending data for Si$_2$C, NCNCS, HNC,
CH$_3$NCO, $^{37}$ClCNO, and OCCCO.

Notwithstanding model Hamiltonian  (\ref{modham}) has the basic ingredients to
model the limiting linear and bent cases, as well as the rich gamut of
intermediate situations, it is too simple  to attain experimental accuracy in fits of observed band origins for
bending degrees of freedom. Previous fits have been performed, in most cases, using the general one- and two-body
algebraic Hamiltonian \cite{Iachello2003,Larese2011,Larese2013}; adding higher
order interactions for specially hard 
cases, as in the case of the bending vibrational spectrum of water \cite{Larese2013}. We have recently
presented improved results from a systematic study using the most general
Hamiltonian that includes  up to four-body interactions \cite{KRivera2020}. 
%\begin{widetext}
  \begin{align}
    \hat H_{4b} =& P_{11} \hat n \nonumber\\
                 & + P_{21} \hat n^2 + P_{22} \hat \ell^2 + P_{23} \hat W^2 \nonumber\\
                 & + P_{31} \hat n^3 + P_{32} \hat n \hat \ell^2 + P_{33} (\hat n \hat W^2 + \hat W^2 \hat n) \label{H4b}\\
                 & + P_{41} \hat n^4 + P_{42} \hat n^2 \hat \ell^2 + P_{43} \hat \ell^4 + P_{44} \hat \ell^2 \hat W^2 \nonumber\\
                 & + P_{45} (\hat n^2 \hat W^2 + \hat W^2 \hat n^2) + P_{46} \hat W^4
                   + P_{47} (\hat {W}^2 \hat {\overline{W}}^2 + \hat {\overline{W}}^2
                   \hat W^2)/2~.\nonumber
  \end{align}
%\end{widetext}
The notation for the algebraic spectroscopic parameters, \(P_{ij}\),
indicates that this is the \(j-th\) parameter for \(i\)-th body
interactions.  The matrix elements of the different operators in
Hamiltonian (\ref{H4b}) in the two possible basis --\(U(2)\) and
\(SO(3)\)-- can be found in Ref.~\cite{KRivera2020}, where the authors
have recently published a fit to bending data of Si$_2$C, NCNCS, and
HNC using Hamiltonian (\ref{H4b}), obtaining a very satisfactory
agreement with the reported band origins. The interpretation of the PR
for the resulting eigenstates \cite{KRivera2020}, is hampered by the
new ESQPT features introduced by three- and four-body interactions in
Hamiltonian (\ref{H4b}) as it was already shown, for a simpler case, in
Ref.~\cite{PBernal2010}. This provides further support to the use of a
basis-independent alternative quantity as the QFS.

In the present work, we carry out similar fits to a selected bending mode of the
CH$_3$NCO, $^{37}$ClCNO, and OCCCO molecules. Though the influence of all
spectroscopic parameters in (\ref{H4b}) was explored for each case under study,
not all of them are
needed in the fit and you can find a summary of our results in Tab.~\ref{tab-params}. For the sake of
brevity, we explain it in detail the results for Si$_2$C, whereas the
results obtained for the rest of the molecules are  more succinctly
reported. Some extra details can be found in the figures included in Appendix. 

In the Hamiltonian  (\ref{H4b})  the  three- and four-body
operators \(\hat
n \hat W^2 + \hat W^2 \hat n\), \(\hat n^2 \hat W^2 + \hat W^2 \hat n^2\), and \((\hat {W}^2 \hat {\overline{W}}^2 + \hat {\overline{W}}^2
                   \hat W^2)/2\) are built as symmetrized products of Casimir
                   operators and, therefore, are not diagonal
                   neither in the \(U(2)\) nor the \(SO(3)\) bases. To take this fact into account, we extend the
                   definitions (\ref{Hlamb},\ref{HI}) to include such operators 
\begin{equation}
  \hat{H} \left(\lambda\right) =  \left(1-\lambda\right) \hat{H}_{I} +
                                   \left(1+\lambda\right) \hat{H}_{II} +
                                   \left(1-\lambda^2\right)\hat{H}_{\text{mix}}
                                                  + \hat{H}_{I-II}~,\label{Hlamb_mix}
%  = &   \hat{H}(\lambda=0)  + \lambda \hat{H}^I \nonumber~,\\
%  \hat{H}^I         = &                       -\hat{H}_{I} +\hat{H}_{II} -2 \lambda \hat{H}_{\text{mix}}~,\label{HI_mix}
\end{equation}
\noindent where $\hat{H}_{I}$, $\hat{H}_{II}$, and $\hat{H}_{I-II}$ have the
same meaning explained in Eq.~(\ref{Hlamb},\ref{HI}), and $\hat{H}_{\text{mix}}$
encompasses those interactions that are diagonal in neither the \(U(2)\)  nor the \(SO(3)\) basis. In this case, and applying first order perturbation theory, the
interaction Hamiltonian is
\(\hat{H}^I = -\hat{H}_{I} +\hat{H}_{II} -2 \lambda \hat{H}_{\text{mix}}\).
Again, the original Hamiltonian is recovered for $\lambda =0$ and the
Hamiltonian \( \hat{H} \left(\lambda=\pm 1\right)\) is diagonal in the \(U(2)\)/\(SO(3)\)  basis. Considering this definition, the QFS can be computed using
Eq.~(\ref{fidsuscepII}).

\subsection{Detailed study of the  Si$_2$C case}
% Experimental data: 37
% N = 49
% rms=1.48
The available data for the large amplitude bending degree of freedom
of Si$_2$C \cite{Reilly2015} were studied using the four-body 2DVM
Hamiltonian (\ref{H4b}), obtaining already a fit within experimental
accuracy considering one- and two-body operators (fitting
spectroscopic parameters $P_{11}$, $P_{21}$, $P_{22}$, and
$P_{23}$). The number of available observed term values is 37, with
$\nu_b$ up to $13$ and a maximum vibrational angular momentum $\ell =
3$~\cite{Reilly2015}. The resulting fit has {\em
  rms}=$1.48~\text{cm}^{-1}$ \cite{KRivera2020}, of the same order
than the reported experimental uncertainty
($2.0~\text{cm}^{-1}$)~\cite{Reilly2015}. The values of the total
vibron number, \(N\), the optimized spectroscopic parameters and their
one-sigma uncertainty, and the fit \(rms\) are shown in
Tab.~\ref{tab-params}. The interested reader can find a detailed
description of the fitting procedure in Ref.~\cite{KRivera2020}.

In this case, as all operators are diagonal in either the  \(U(2)\)  or the \(SO(3)\) basis (\ref{U3chains}), the control parameter \(\lambda\) dependent
Hamiltonian (\ref{Hlamb}) can be written as follows

%\begin{align}
\begin{equation}
  \begin{matrix}
  \hat{H}_{\text{Si}_2\text{C}}\left(\lambda\right) &=&\left(1-\lambda\right)&\left[P_{11} \hat{n} + P_{21} \hat{n}^2  \right]  &+& \left(1+\lambda\right) & \left[ P_{23}\hat{W}^2 \right] &+&\left[P_{22}\hat{\ell}^2\right] \nonumber\\
  &=& \left(1-\lambda\right) & \hat H_{I} &+& \left(1+\lambda\right)& \hat H_{II} &+& \hat H_{I-II} \nonumber
  \end{matrix}
\end{equation}
%\end{align}

The optimized Hamiltonian parameter values are obtained for $\lambda = 0$, as
\(\lambda\) approaches a value of \(1\) (\(-1\)) only terms associated with
the \(SO(3)\) (\(U(2)\)) dynamical symmetry are nonzero.

\begin{figure} \centering \includegraphics[trim = 100 150 0 180,
  clip, width=1.0\textwidth]{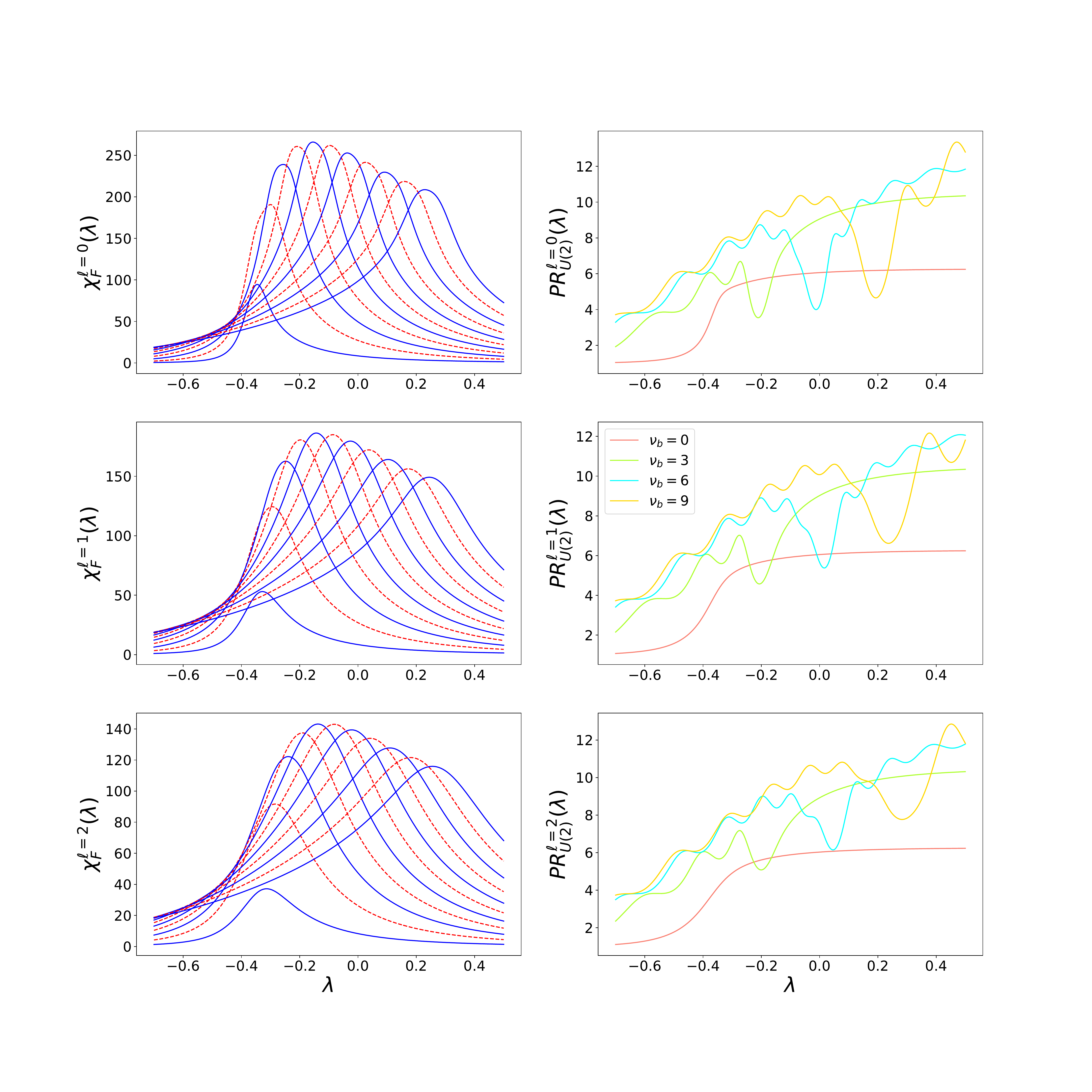}
  \caption{\label{Si2C_vs_lamb} All panels: Results for the optimized
    eigenstates of the bending degree of freedom of Si$_2$C. The upper, middle,
    and lower rows show results for \(\ell = 0,1,2\), respectively. Left panels:
    QFS for states with $\nu_b = 0,1,\ldots,10$ as a
    function of the \(\lambda\) control parameter. Full blue lines alternate
    with dashed red lines for an easier distinction between adjacent states.
    Right panels: Participation ratio versus the $\lambda$ control parameter for
    selected Si$_2$C bending states (\(\nu_b = 0, 3, 6\), and \(9\)). }
\end{figure}

We proceed to calculate the QFS as a function of
$\lambda$ for the first eleven Si$_2$C bending eigenstates, well beyond the
barrier to linearity. The obtained results,
for vibrational angular momentum $\ell = 0,1,$ and $2$, are depicted in the left
column panels of Fig.~\ref{Si2C_vs_lamb}. For each state, the QFS is maximal at a certain \(\lambda\) value. This result is
completely equivalent to the result presented in Fig.~\ref{mHsus} for the model
Hamiltonian (\ref{modham}): the maximum QFS value indicates what is the
$\lambda$ value for which the state under study crosses the high-density of states ESQPT
separatrix line. Therefore, if the maximum occurs for a negative \(\lambda\)
value, an originally bent  state (belonging to the $SO(3)$ or broken symmetry phase) is
changing to a linear  state
(that belongs to the $U(2)$ or symmetric phase) and vice versa for a maximum at a positive \(\lambda\)
value. In nonrigid molecules, where an ESQPT is expected, the level whose  QFS maximum
is the closest to $\lambda = 0$ is the bending eigenstate with an energy that
is equal to the ESQPT critical energy. This state can be considered the transition state from
bent to linear configurations. The states of the left column of
  Fig.~\ref{Si2C_vs_lamb} can be labeled attending to their maximum position:
  from left to right we have included states $\nu_b=0,1,...,10$ and, in the
  Si\(_2\)C case, the transition state is the fifth bending overtone,  $\nu_b
  = 6$.

The right column panels in Fig.~\ref{Si2C_vs_lamb} show
the participation ratio values in the $U(2)$ basis for bending levels
$\nu_b=0,3,6$, and $9$, and for vibrational angular momentum values $\ell=0,1$,
and $2$.
As already mentioned in
the discussion of the lower panel of Fig.~\ref{mHsus}, in the calculation of the
participation ratio for the model Hamiltonian, the system  ground state is
better localized in the $U(2)$
basis before the ground state QPT, and the PR value suddenly increases once the system goes through the critical
point. As this sudden change takes place for a negative \(\lambda\) value, that
implies that the ground state of
Si$_2$C is a bent-like state, as expected. In the excited levels
case, the PR is minimal once each wave function gets through the ESQPT separatrix, being $\nu_b=6$ the most localized state for
$\lambda=0$. We have performed also calculations of the QFS for the optimized Si$_2$C
states (\(\lambda = 0\) case) with vibrational angular momentum \(\ell = 0,1\),
and 2. The obtained results are shown in Fig.~\ref{Si2C_lmb0} in Appendix
\ref{AppA}; and they display the same trends than the model Hamiltonian case shown
in the upper panel of Fig.~\ref{mHlmax}.  

As expected, in both cases, for the PR and the QFS, ESQPT precursors
are weaker for higher $\ell$ values (see also Fig.~\ref{Si2C_lmb0} in
Appendix \ref{AppA}). This is a well-known effect explained by the
centrifugal barrier hindering the access of the wavefunction to the
maximum in the barrier to linearity \cite{Caprio2008}.

% The QFS and PR values for the Si$_2$C eigenstates resulting from the fit for
% different vibrational angular momentum values has been provided in the
% Supplementary Material section, with results that are equivalent to the ones
% obtained for the model Hamiltonian in the upper panel of Fig.~\ref{mHlmax}.

%%%%
\subsection{Application to other molecules}
% HNC, CH3NCO, ClCNO, OCCCO, NCNCS
The detailed study of the Si$_2$C case evinces the efficiency of the
QFS locating excited states with respect to the ESQPT
separatrix. In the present subsection, we extend the study to other molecular
species. We include a linear molecule (HNC) and four nonrigid species
(CH$_3$NCO, \(^{37}\)ClCNO, OCCCO, and NCNCS) with different barrier to
linearity heights.

As in the Si$_2$C case, the fits for the HNC and NCNCS are the same than the
authors have recently presented in \cite{KRivera2020}, using the 2DVM four-body Hamiltonian (\ref{H4b}).
In short, the  HNC fit included terms up to two-body interactions plus a
three-body term, \(\hat n \hat\ell^2\) ($P_{32}$ parameter), to fit 19 experimental data~\cite{MellauHNC} with
$N=40$. The obtained fit precision is very satisfactory, with an $rms=0.08\,\text{cm}^{-1}$. The
modeling of the NCNCS nonrigid bending degree of freedom (CNC bending, \(\nu_7\) normal mode)
implies  the inclusion of one- and two-body terms plus two four-body operators
($P_{42}$ and $P_{46}$) to reproduce the $88$ reported bending band origin values~\cite{Winnewisser2010} with
an $rms=0.79\,\text{cm}^{-1}$ ($N=150$).

We have carried out fits,  using the four-body Hamiltonian (\ref{H4b}), to the
available data for nonrigid  bending vibrational modes  for CH$_3$NCO, $^{37}$ClCNO, and OCCCO.  The molecules CH$_{3}$NCO
and $^{37}$ClCNO were previously studied using the 2DVM, but including only
interactions up to 2-body in the Hamiltonian~\cite{Larese2011}. In
all cases, the one- and two-body interactions have been included. In the
CH$_3$NCO case we have added an additional (four-body) term, in the
$^{37}$ClCNO case two additional (three-body and four-body) terms,  and one
extra (four-body) term in the OCCCO case.

We show in Tab.~\ref{tab-params}
the optimized parameter values for the different molecules under study. For the sake of
completeness we include the parameter values for the six molecular species. In
all cases $P_{ij}$ coefficients are reported in
cm$^{-1}$ units and we include  the total number of bosons $N$, the achieved root mean square \(rms\) deviation (cm$^{-1}$) and the total number of experimental data
$N_{data}$ used in the fit. A detailed description of the fitting procedure can be
found in Ref.~\cite{KRivera2020}. We include tables with the reported bending
band origins, our calculation and state assignment as well as the fit residuals
in App.~\ref{AppB}.

\begin{table}[H]
  \centering
  \scriptsize
  % \begin{tabular}{c|ccccccc}
  %   \hline
  %   \hline
  %                & CH$_3$NCO                 & $^{37}$ClCNO              & OCCCO                   \\ 
  %   \hline
  %   $P_{11}$     & 449.5(13)                 & 758.73(15)                & 263.99(15)              \\
  %   $P_{21}$     & -5.477(22)                & -7.8895(23)               & -2.3308(25)             \\
  %   $P_{22}$     & 7.85(7)                   & 3.85(4)                   & 1.300(17)               \\
  %   $P_{23}$     & -1.628(4)                 & -2.1217(6)                & -0.6768(4)              \\
  %   $P_{33}$     & -                         & 9(7)$\times 10^ {-6}$     & -                       \\ 
  %   $P_{42}$     & -                         & -3(3)$\times 10^ {-4}$    & -                       \\
  %   $P_{43}$     & -                         & -                         & 7.2(12)$\times 10^{-4}$ \\
  %   $P_{45}$     & -1.25(23)$\times 10^{-5}$ & -6.46(14)$\times 10^{-6}$ & -                       \\
  %   \hline
  %   \(N\)            & 78                        & 92                        & 100                     \\
  %   \hline
  %   {\em rms }   & 1.10                      & 0.095                     & 0.60                    \\
  %   \(N_{data}\) & 19                        & 33                        & 36                      \\
  %   \hline
  %   \hline
  % \end{tabular}

    \begin{tabular}{c|cccccccccc}
    \hline
    \hline
                 & CH$_3$NCO                 & $^{37}$ClCNO              & OCCCO                  & HNC$^a$                     &Si$_2$C$^a$   & NCNCS$^a$ \\ 
    \hline
    $P_{11}$     & 449.5(13)                 &  760.88(16)               & 263.99(15)             &  1414.0(4)                  & 63.8(5)     & 331.97(8) \\
    $P_{21}$     & -5.477(22)                & -7.9142(24)               & -2.3308(25)            & -29.837(15)                 & -0.108(18)  & -2.0954(6)\\
    $P_{22}$     & 7.85(7)                   &  3.818(14)                & 1.300(17)              &  15.81(10)                  & 0.98(5)     &1.190(8) \\
    $P_{23}$     & -1.628(4)                 &  -2.1276(6)               & -0.6768(4)             &  -8.054(3)                  & -0.8117(17) &-0.58578(17)\\
    $P_{32}$     & -                         & -                         & -                      &   4.9(10) $\times 10^{-2}$   & -           &- \\
    $P_{33}$     & -                         &  1.8(7)$\times 10^{-5}$    & -                      & -                           & -           &-\\ 
    $P_{42}$     & -                         & -                         & -                      & -                           & -           & -2.65(20)$\times 10^{-5}$\\
    $P_{43}$     & -                         &  -                        & 7.2(12)$\times 10^{-4}$ & -                           & -           &- \\
    $P_{45}$     & -1.25(23)$\times 10^{-5}$ & -6.61(15)$\times 10^{-6}$  & -                       & -                           & -           &-\\
    $P_{46}$     & -                         & -                         & -                      & -                           & -           & 3.48(8)$\times 10^{-7}$\\
    \hline
    \(N\)            & 78                   & 92                        & 100                     & 40                          & 49          &150 \\
    \hline
    {\em rms }   & 1.10                      & 0.12                     & 0.60                    & 0.08                        & 1.48        & 0.79 \\
    \(N_{data}\) & 19                        & 33                        & 36                      & 19                          &  37         & 88 \\
    \hline
    \hline
  \end{tabular}
  \caption{Optimized Hamiltonian parameters ($P_{ij}$, in cm$^{-1}$
    units) for the selected bending degree of freedom of the molecules under study. Values are
    provided together with their associated uncertainty in parentheses in units of
    the last quoted digits. The total vibron number, \(N\), the obtained {\em
      rms} of the fit, and the number of reported bending band origins considered in the
    fit are also included.}
  \begin{flushleft}
    $^a$ Fits from a previous work \cite{KRivera2020}. 
  \end{flushleft}
  \label{tab-params}
\end{table}

% Brief revision of each new molecule
% 
% CH3NCO
% % Exp: vb_max=3, elle_max=7
% % rms Danielle: 1.34
The CNC bending of CH$_{3}$NCO (normal mode $\nu_{8}$) has a nonrigid
character and we have carried out a fit  making use of the
four-body Hamiltonian (\ref{H4b}) to the  19
available experimental data~\cite{Koput1986}, with $\nu_b$ up to 3, and
a maximum value of the vibrational angular momentum $\ell=7$. The parameter resulting from the fit can be
found in the first column of Tab.~\ref{tab-params}. The obtained results,
with an $rms = 1.10$~cm$^{-1}$, are rather
close to the results previously obtained with the 2DVM including only one- and
two-body interactions in the Hamiltonian ($rms = 1.34$~cm$^{-1}$)
\cite{Larese2011}. We have kept constant the total number of
bosons $N$ used in Ref.~\cite{Larese2011} and there is only a four-body
parameter from  Hamiltonian (\ref{H4b}), $P_{45}$, that significantly improves
the quality of the fit. This can be explained due to the complexity of the
CH$_{3}$NCO spectrum, with two coupled vibrational modes
of large amplitude: an internal methyl rotor, with a low energy potential barrier
(at approx.~$20$~cm$^{-1}$), and the CNC bending mode, characterized by a large
anharmonicity. This molecule is currently the target of some studies in our
group, trying to simultaneously treat the large amplitude bending and the
internal rotation within a common algebraic formalism. 

% 37ClCNO
% % Exp: vb_max=3, elle_max=9
% % rms Danielle: 0.71
We have also performed a fit to the 33 available experimental data  for the ClCN
bending ($\nu_5$ normal mode) of the $^{37}$Cl
isotopologue of ClCNO~\cite{Br_Cl_CNO2001}. The data set comprises states with
bending excitation $\nu_b$ up to
3 and vibrational angular momentum $\ell$ up to 9 units. As in the CH$_3$NCO
case, the bending spectrum of this normal mode has been  
previously analyzed using the one- and two-body Hamiltonian of the 2DVM,
obtaining  an {\em rms} of
$0.71\text{cm}^{-1}$ ~\cite{Larese2011}. Our fit includes two
higher-order interactions: $P_{33}$ and $P_{45}$, which allows for a
a reduction of the $rms$ to $0.12\,\text{cm}^{-1}$. In this case, the
interactions introduced are  diagonal in neither the \(U(2)\) nor the \(SO(3)\) basis. As in the previous case,  we have used the same
total number of vibrons, $N$, than Larese \textit{et al.}~\cite{Larese2011}.

% OCCCO
% % Exp: vb_max=3, elle_max=12
The third molecular species whose bending spectrum has been modeled for its
inclusion  in the present work is OCCCO. In this case we focus on the CCC
bending (normal mode $\nu_7$) and we have carried out a fit to the 36 available
experimental term values, with a maximum  $\nu_b=3$  and a maximum
$\ell=12$~\cite{OCCCO_2}. We have included, in addition to one- and two-body operators, the  $P_{43}$ parameter interaction, obtaining an \(rms\) equal to
$0.60\,\text{cm}^{-1}$.  The total number of vibrons has been manually adjusted to
$N = 100$.
%% In this case the {\em rms} has a
%% asymptotic behavior as the total number of bosons $N$ grows. Looking
%% for a high quality fit without spending too much time in calculations,
%% we decided to fix this value to $100$.

%\begin{widetext}
\begin{figure} \centering \includegraphics[trim = 0 0 0 0,
clip, width=1.\textwidth]{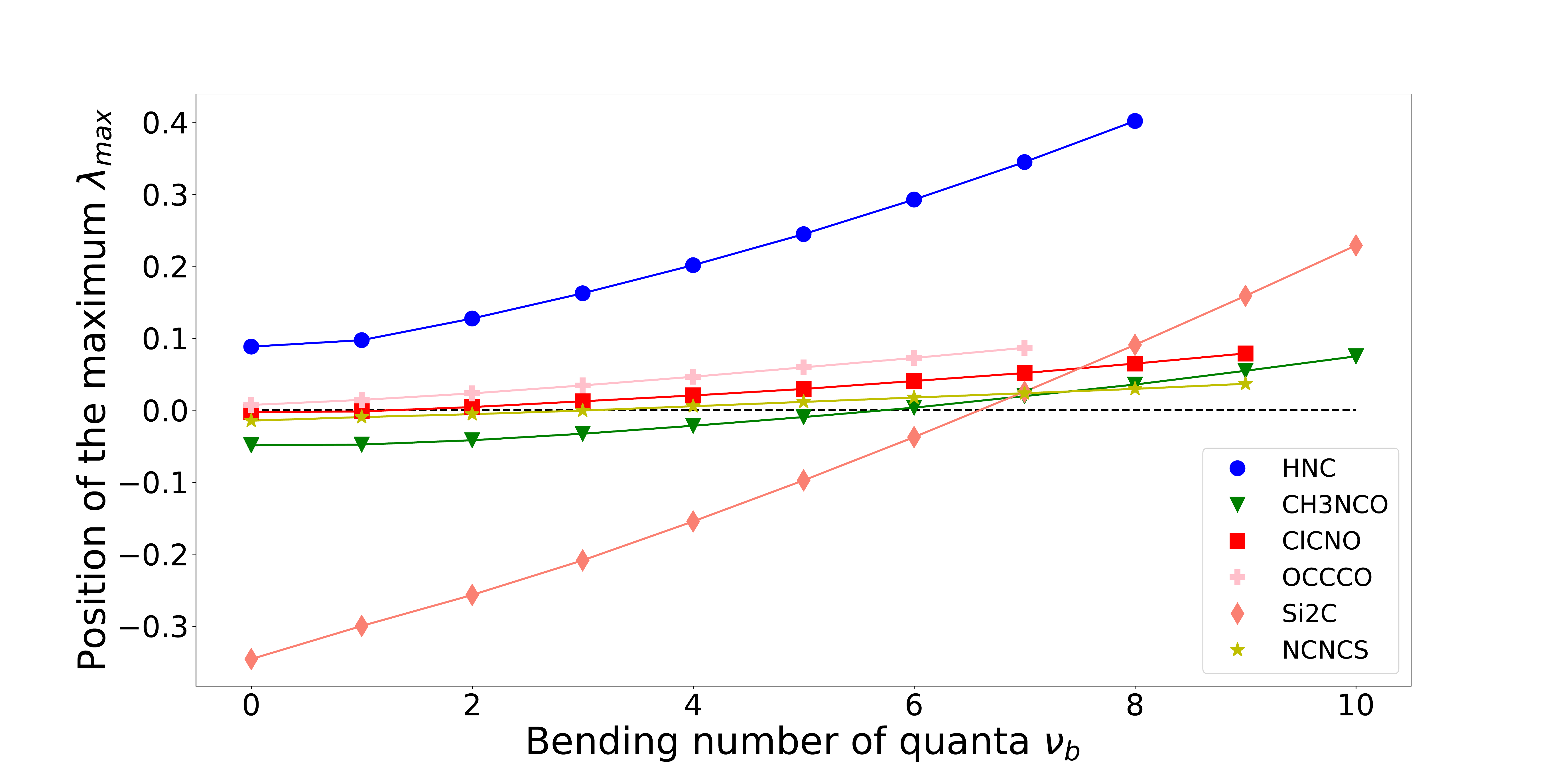}
\caption{\label{lambdas_max} Position $\lambda$ where the
  QFS takes its maximum value for each state with $\ell=0$
  of selected bending degrees of freedom for HNC (blue), CH$_3$NCO (dark-green), $^{37}$ClCNO (red), OCCCO
  (pink), Si$_2$C (coral) and NCNCS (olive), versus the bending
  quantum number $\nu_b$. The dashed black horizontal line marks the \(\lambda = 0\) value.}
\end{figure}
%\end{widetext}
In summary, we have located where the maximum QFS occurs for states with
$\ell=0$. In all cases, the optimized Hamiltonian can be recovered for
$\lambda = 0$. For this reason, attending to the explanation given
in section~\ref{sec3}, a state with a maximum QFS at a negative $\lambda$
value is located in the $SO(3)$ phase of the ESQPT and should have a bent
character. In case the maximum QFS occurs at a positive $\lambda$, the state
has a linear character and belongs to the $U(2)$ ESQPT phase.

The Fig.~\ref{lambdas_max} shows the $\lambda$ values at which maxima occur for
the states with
$\ell=0$ in all cases examined, including Si$_2$C. HNC and Si$_2$C 
can be considered as two textbook examples of a linear molecule and of a
nonrigid molecule. As
expected, all $\lambda$ values are positive in the HNC case (blue circles). The varying slope
between the ground and the first excited states and the rest could be explained by the change from a
positive to a negative anharmonicity that characterizes the bending of this molecular species. On the other
hand, in the Si$_2$C case (coral diamonds), the ground state, the fundamental, and the first five overtones have
bent character (negative $\lambda$), whereas the rest are linear.

The results for CH$_3$NCO (green triangles), $^{37}$ClCNO (red squares), and NCNCS (olive stars)
confirm that these molecules are also
nonrigid, and their excited states lie closer to the separatrix line between
the ESQPT phases than in the  Si$_2$C case. Therefore, their QFS maxima
occur in the vicinity of $\lambda=0$. The states with QFS maximum at negative
$\lambda$ values are the ground state plus five excited
states for CH$_3$NCO  and the ground state plus two excited states for
NCNCS. Therefore, the following excited state  would be just above the barrier to linearity.
In the \(^{37}\)ClCNO isotopologue case, already
the bending fundamental has a maximum at a positive $\lambda$ value and, therefore,
only the ground state with $K_a=J=0$ would have bent character. The last molecule we have decided to include in this
work is OCCCO (pink crosses), with all states above a very low energy barrier to
linearity, including the ground state. 

We have included in the left column of Fig.~\ref{MolCom} of App.~\ref{AppB} the dependence of
the QFS with $\lambda$ for the above mentioned five molecules, from where the
position of the maxima reported in Fig.~\ref{lambdas_max} have been
extracted. In the right column panels of the same figure, we depict the PR in the two bases
considered as well as the \(\lambda=0\) QFS values.

\section{Concluding remarks}
\label{sec5}

In summary, we have introduced a new perspective into excited state quantum
phase transitions making use of the quantum fidelity susceptibility in the study
of the excited states of the 2D limit of the vibron model. The QFS, a quantity
of first importance in Quantum Information Theory, has been chiefly used to
characterize ground state quantum phase transitions in different many-body
quantum systems. Using results for a 2DVM model Hamiltonian, we have shown how the
extension of the QFS from the ground state to encompass excited states provides
a convenient tool for the study and characterization of ESQPTs and allows for a
fully basis-independent assignment of overtones to one of the possible ESQPT
phases in molecular bending spectra. In this regard, these findings nicely complements the information about
the ESQPT provided by the PR \cite{Santos2015, Santos2016, PBernal2017}, though
QFS achieves an unambiguous assignment of states to ESQPT phases even for
situations very far from the dynamical symmetry limits.

As an application, we have carried out calculations using a four-body algebraic
2DVM Hamiltonian (\ref{H4b}) and fitting bending data from six molecular
species: Si$_2$C, HNC, CH$_3$NCO, $^{37}$ClCNO, OCCCO, and NCNCS. The fits to
reported band origins of three of them (Si$_2$C, HNC, and NCNCS) have been
recently published \cite{KRivera2020}, while the fits for the other three
(CH$_3$NCO, $^{37}$ClCNO, OCCCO) have not been previously reported. In all
cases, the vibrational bending mode under study is anharmonic, and all but
HNC can be considered as nonrigid molecular species, with a feature-rich and
complex bending spectrum. A very satisfactory agreement with the
reported data has been achieved; the obtained energies and eigenfunctions have
been used for the calculation of QFS for the six molecular species.

We have presented a detailed account of the QFS results obtained in
the Si$_2$C case, and an outline of the results for the rest of the
molecules. The obtained results provide a satisfactory estimation of the
height of the barrier to linearity (which coincides with the ESQPT
critical energy) in all cases and the QFS has proved to be a very
sensitive tool for the classification of eigenstates as having a
linear or bent character.

We are currently working on the study of universality and scaling laws of the
QFS in ground state and excited state QPTs for the vibron model and its limits
in 1D and 3D as well as in ESQPTs for other quantum systems.

\section*{Acknowledgements}
We thank useful discussion with Profs.~Lea F.~Santos, José M.~Arias, and
Pedro Pérez-Fernández.  Computing
resources supporting this work were provided by the CEAFMC and Universidad de
Huelva High Performance Computer (HPC@UHU) located in the Campus Universitario
el Carmen and funded by FEDER/MINECO project UNHU-15CE-2848.

% % TODO: include author contributions
% \paragraph{Author contributions}
% This is optional. If desired, contributions should be succinctly described in a single short paragraph, using author initials.

% TODO: include funding information
\paragraph{Funding information}

This project has received funding from the European Union's
Horizon 2020 research and innovation program under the Marie Sk\l odowska-Curie
grant agreement No 872081 and from the Spanish National Research, Development,
and Innovation plan (RDI plan) under the project PID2019-104002GB-C21 (JKR, MC,
and FPB) and COOPB20364 (MC).  This
work has also been partially supported by the Consejer\'{\i}a de Conocimiento,
Investigaci\'on y Universidad, Junta de Andaluc\'{\i}a and European Regional
Development Fund (ERDF), refs.  SOMM17/6105/UGR (MC and FPB) and UHU-1262561 (JKR and FPB). 

\begin{appendix}

% \section{First appendix}
% Add material which is better left outside the main text in a series of Appendices labeled by capital letters.

  \section{Centrifugal barrier effects}
  \label{AppA}

\begin{figure}[H]
    \centering
    \includegraphics[trim = 0 100 0 180, clip,
    width=0.6\textwidth]{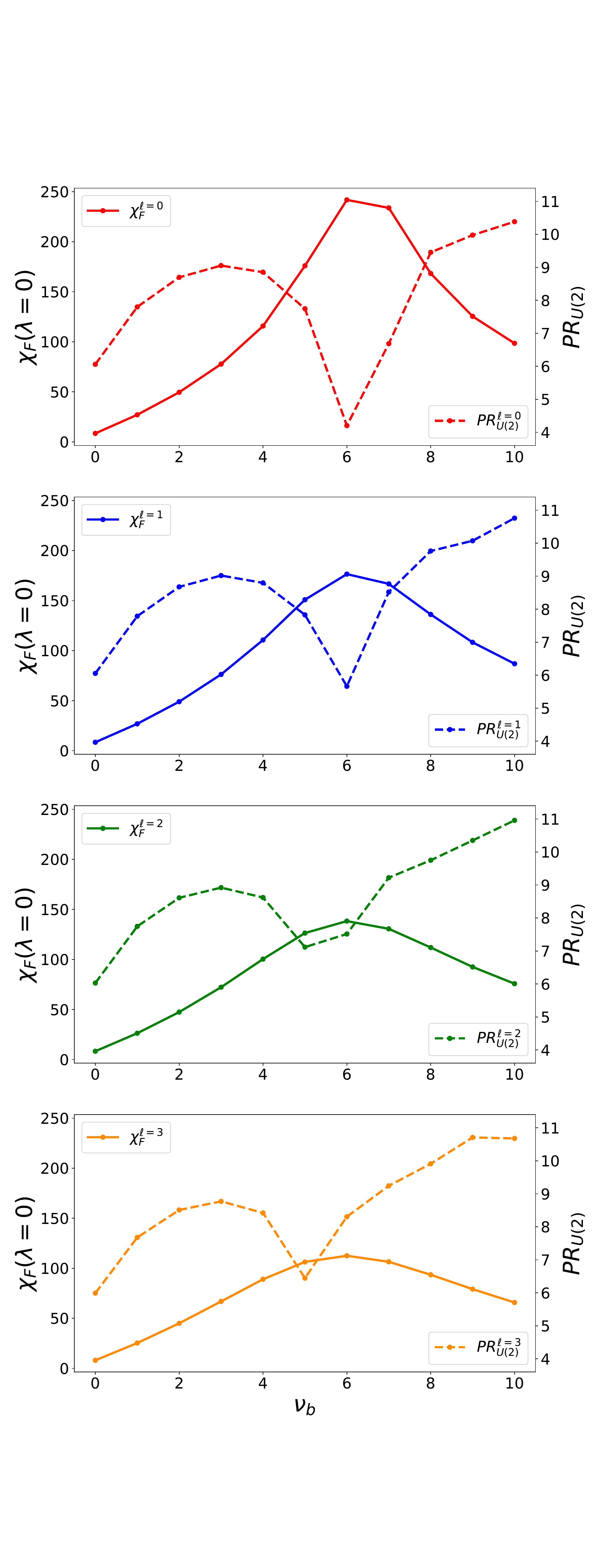}
    \caption{ Si$_2$C bending eigenstates QFS (solid line) and PR in
      the $U(2)$ basis (dashed line) evaluated for $\lambda = 0$
      versus the bending quantum number $\nu_b$. Red, blue, green, and
      orange lines correspond to $\ell=0$, $1$, $2$, and $3$,
      respectively.}
    \label{Si2C_lmb0}
\end{figure}

  We illustrate the centrifugal barrier effects over QFS and PR with
  the data calculated for the Si\(_2\)C molecule in
  Fig.~\ref{Si2C_lmb0}. We show the QFS (full lines and left ordinate
  axes) and the PR in the \(U(2)\) basis (dashed lines and right ordinate
  axes) for vibrational angular momentum values $\ell=0$ (red, first panel), $1$
  (blue, second panel), $2$ (green, third panel), and $3$ (orange, fourth panel). These two quantities are excellent probes
  to look for ESQPT precursors in the 2DVM and other systems. In the
  panels of this figure, we can appreciate how the
  bent-to-linear ESQPT precursors weaken for increasing vibrational
  angular momentum values.

  It is known that the ESQPT critical state in bent-to-linear transitions modeled with the 2DVM has a large
  component in the $\ket{n^{\ell}}=\ket{\ell^{\ell}}$ element when expressed in the $U(2)$ basis \cite{Santos2015,Santos2016,PBernal2017}, which translates  into a
  minimum value of the Participation Ratio. The localization in the U(2)
  basis becomes softer for higher values of $\ell$. This well known fact can be explained
  considering the influence of the centrifugal barrier, which hinders
  the non-zero angular momentum wave function access to the
  bent-to-linear barrier maximum.
 
  It can be appreciated in the figure how QFS values $\chi_F\left(\lambda=0\right)$ in the
  transition state $\nu_b=6$ diminish as the
  vibrational angular momentum $\ell$ increases.

  \section{Energy fits, residuals, and QFS and PR results}
  \label{AppB}

  As already mentioned, the procedure followed to fit the Hamiltonian
  (\ref{H4b}) to the available bending origin bands is the one already
  described in Ref.~\cite{KRivera2020}. For the sake of completeness,
  we include in this appendix tables including experimental and
  computed bending band origin values, as well as the resulting
  residuals. We provide the results for the
  three species whose fit has not been published yet: CH$_3$NCO
  (Tab.~\ref{resCH3NCO}), $^{37}$ClCNO (Tab.~\ref{resClCNO}), and
  OCCCO (Tab.~\ref{resOCCCO}).  The states are labeled in all cases but the OCCCO one
  using the bent molecule notation. In the OCCCO case we use the
  linear molecule quantum labels transforming them to \(\nu_b\) values
  in the figures \cite{KRivera2020}. The interested reader can find results for
  HNC, Si$_2$C, or NCNCS in Ref.~\cite{KRivera2020}.

\begin{figure}[H]
    \centering \includegraphics[trim = 0 150 0 200, clip,
    width=0.6\textwidth]{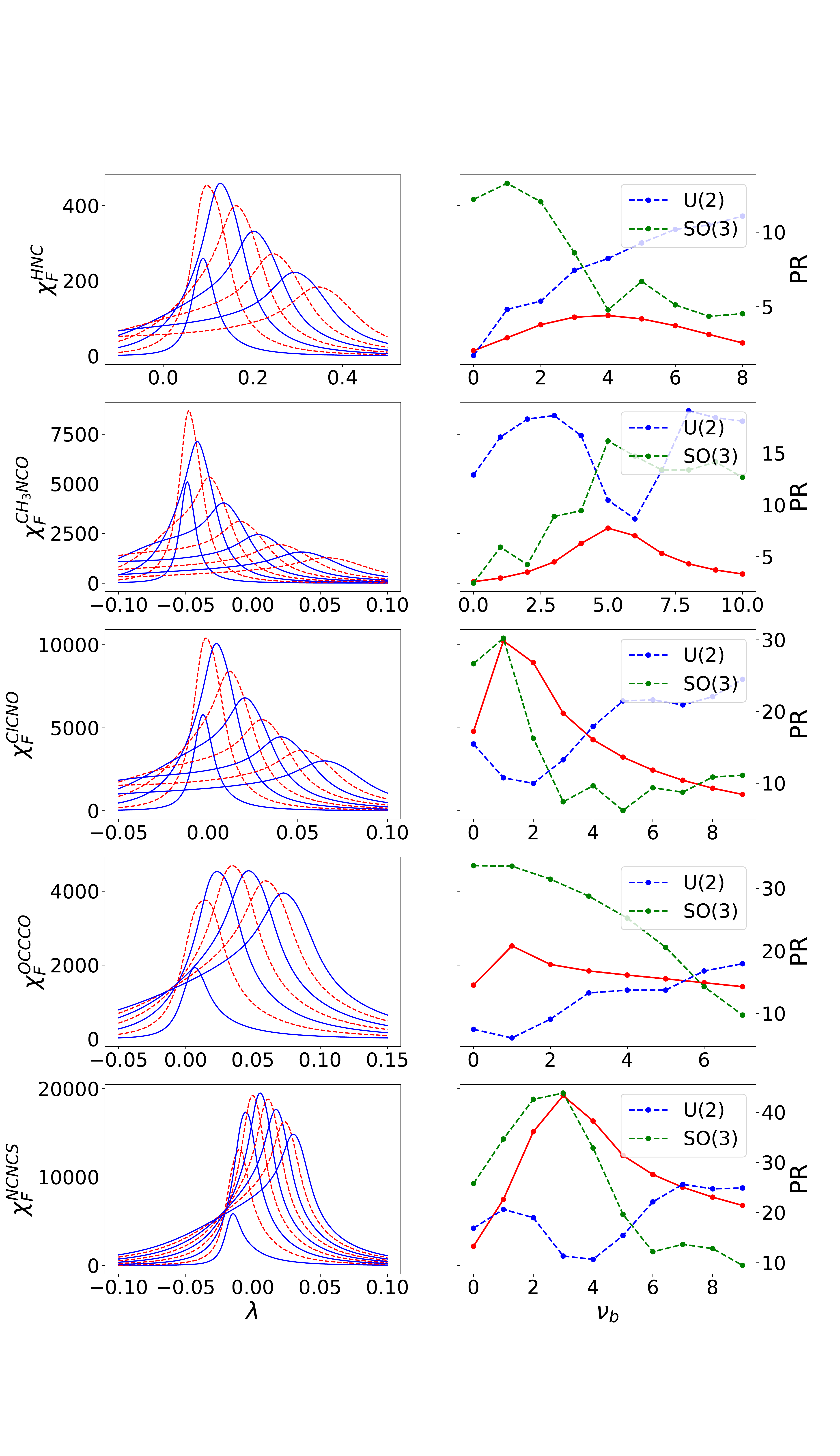}
    \caption{Left column: QFS for states
      with $\ell=0$, \(\chi_F(\lambda)\), versus the control parameter, $\lambda$. Right column: QFS for \(\lambda =0\) (solid red line) using the same scales as in the left panels and  PR in the $U(2)$ (blue dashed
      line) and $SO(3)$ (green dashed line) bases (right axes scale). Results for the five
      molecules that have been selected to illustrate the QFS results in 2DVM systems, from top to bottom: HNC, CH$_3$NCO, $^{37}$ClCNO,
      OCCCO, and NCNCS.}
    \label{MolCom} 
  \end{figure}

We also provide in this appendix the intermediate results needed to
reproduce Fig.~\ref{lambdas_max}. In the left column of
Fig.~\ref{MolCom}, the QFS for the first bending states is depicted as
a function of the \(\lambda\) parameter. QFS values for \(\nu_b = 0,
2,\ldots\) states are depicted with full blue lines, while for \(\nu_b
= 1, 3,\ldots\), the QFS is depicted with dashed red lines.  The
\(\lambda\) parameter values corresponding to the maximum QFS value
for each state in these panels are the ones depicted in
Fig.~\ref{lambdas_max}.

The panels in the right column of Fig.~\ref{MolCom} display the
$\lambda=0$ QFS (\(\chi_F(\lambda=0)\), red full lines), and PR in the
U(2) (blue dashed lines) and SO(3) (green dashed lines) bases as a
function of the number of quanta of bending excitation for each
molecule. The QFS shares axes and ticks with the corresponding plot on
the left column, whereas the scale for the PR is located on the right
side of the right column panels.
  
  \begin{table}[H]
    \centering %\scriptsize
\begin{minipage}{0.4\textwidth}
  \begin{tabular}{cccc}
    \hline
    $\nu_b$,~ $\ell$ &       Exp. E.  &       Calc. E. & Exp.-Calc. \\
    \hline
    0,~  0  &        0.0  &      0.0000    &      0.0000\\
    1,~  0  &      182.2  &    183.5477    &     -1.3478\\
    2,~  0  &      357.9  &    358.8119    &     -0.9119\\
    3,~  0  &      525.1  &    523.8246    &    1.2754  \\
    0,~  1  &        8.4  &      8.7997    &     -0.3997\\
    1,~  1  &      191.4  &    193.0995    &     -1.6995\\
    2,~  1  &      368.6  &    369.6086    &     -1.0086\\
    0,~  2  &       36.8  &     35.1316    &    1.6684  \\
    1,~  2  &      222.3  &    221.5793    &    0.7207  \\
    2,~  2  &      402.1  &    401.4326    &    0.6674  \\
    \hline
  \end{tabular}
\end{minipage}
\hfill\vline\hfill
\begin{minipage}{0.4\textwidth}
  \begin{tabular}{cccc}
    \hline
    $\nu_b$,~ $\ell$ &        Exp. E. &      Calc. E. & Exp.-Calc. \\
    \hline
    0,~  3  &       80.0  &     78.8075    &    1.1925  \\
    1,~  3  &      268.6  &    268.5326    &    0.0674  \\
    2,~  3  &      454.0  &    453.1013    &    0.8987  \\
    0,~  4  &      140.6  &    139.5489    &    1.0511  \\
    1,~  4  &      333.4  &    333.3653    &    0.0347  \\
    0,~  5  &      217.5  &    217.0241    &    0.4759  \\
    1,~  5  &      415.5  &    415.4570    &    0.0430  \\
    0,~  6  &      311.1  &    310.8790    &    0.2210  \\
    1,~  6  &      513.4  &    514.2226    &     -0.8226\\
    0,~  7  &      420.0  &    420.7589    &     -0.7589\\
    \hline
  \end{tabular}
\end{minipage}

    \caption{Experimental~\cite{Koput1986} and calculated band origins and residuals for the CNC bending mode of
      CH$_3$NCO. Units of cm$^{-1}$.}
    \label{resCH3NCO}
  \end{table}

  \begin{table}[H]
    \centering
    %\scriptsize
\begin{minipage}{0.4\textwidth}
  \begin{tabular}{cccc}
    \hline
    $\nu_b$,~ $\ell$ &       Exp. E.  &       Calc. E. & Exp.-Calc. \\
    \hline
    0,~  0 &       0.0     &       0.0000  &    0.0000 \\
    1,~  0 &     120.9     &     120.8932  &    0.0068   \\ 
    2,~  0 &     258.5     &     258.6117  &     -0.1117 \\
    3,~  0 &     432.0     &     432.1241  &     -0.1241 \\
    0,~  1 &      17.5     &      17.6438  &     -0.1438 \\
    1,~  1 &     167.9     &     167.7928  &    0.1072   \\
    2,~  1 &     335.1     &     335.0671  &    0.0329   \\
    3,~  1 &     525.3     &     525.4485  &     -0.1485 \\
    0,~  2 &      55.6     &      55.7618  &     -0.1618 \\
    1,~  2 &     227.8     &     227.7131  &    0.0869   \\
    2,~  2 &     415.1     &     415.1037  &     -0.0037 \\
    3,~  2 &     620.1     &     620.1478  &     -0.0478 \\
    0,~  3 &     108.1     &     108.2590  &     -0.1590 \\
    1,~  3 &     297.6     &     297.4773  &    0.1227   \\
    2,~  3 &     500.5     &     500.4730  &    0.0270   \\
    3,~  3 &     717.9     &     717.9603  &     -0.0603 \\
    0,~  4 &     171.8     &     171.9164  &     -0.1165 \\
    \hline
  \end{tabular}
\end{minipage}
\hfill\vline\hfill
\begin{minipage}{0.4\textwidth}
  \begin{tabular}{cccc}
    \hline
    $\nu_b$,~ $\ell$ &        Exp. E. &      Calc. E. & Exp.-Calc. \\
    \hline
    1,~  4 &     375.5     &     375.4203  &    0.0797   \\
    2,~  4 &     591.3     &     591.2717  &    0.0283   \\
    3,~  4 &     819.6     &     819.4564  &    0.1436   \\
    0,~  5 &     244.7     &     244.7522  &     -0.0522 \\
    1,~  5 &     460.5     &     460.4442  &    0.0558   \\
    2,~  5 &     687.2     &     687.2587  &     -0.0588 \\
    3,~  5 &     925.0     &     924.7723  &    0.2277  \\
    0,~  6 &     325.4     &     325.4264  &     -0.0264 \\
    1,~  6 &     551.8     &     551.7516  &    0.0484  \\
    2,~  6 &     788.1     &     788.1261  &     -0.0261 \\
    0,~  7 &     413.1     &     412.9750  &    0.1250   \\
    1,~  7 &     648.7     &     648.7310  &     -0.0310 \\
    2,~  7 &     893.5     &     893.5698  &     -0.0698 \\
    0,~  8 &     506.8     &     506.6723  &    0.1277  \\
    1,~  8 &     750.8     &     750.8960  &     -0.0960 \\
    0,~  9 &     606.1     &     605.9530  &    0.1470  \\
    1,~  9 &     857.6     &     857.8490  &     -0.2490 \\
    \hline
  \end{tabular}
\end{minipage}

    \caption{Experimental~\cite{Br_Cl_CNO2001} and calculated band origins and residuals for the ClCN bending mode of
      $^{37}$ClCNO. Units of cm$^{-1}$.}
    \label{resClCNO}
  \end{table}

  \begin{table}[H]
    \centering
    %\scriptsize
\begin{minipage}{0.4\textwidth}
  \begin{tabular}{cccc}
    \hline
    $n^{\ell}$ &       Exp. E.  &       Calc. E. & Exp.-Calc. \\
    \hline
    $  2 ^ 0$  &      60.70    &     60.2640    &       0.4360   \\
    $  4 ^ 0$  &     144.30    &    144.3234    &       -0.0234  \\
    $  6 ^ 0$  &     244.70    &    244.4930    &       0.2070   \\
    $  1 ^ 1$  &      18.26    &     18.6991    &       -0.4391  \\
    $  3 ^ 1$  &      97.22    &     97.2062    &       0.0138   \\
    $  5 ^ 1$  &     191.06    &    191.4044    &       -0.3444  \\
    $  7 ^ 1$  &     299.26    &    298.0710    &       1.1890   \\
    $  2 ^ 2$  &      46.11    &     46.5075    &       -0.3975  \\
    $  4 ^ 2$  &     137.26    &    137.4453    &        -0.1853 \\
    $  6 ^ 2$  &     239.57    &    240.1829    &        -0.6129 \\
    $  8 ^ 2$  &     352.91    &    352.8677    &       0.0423   \\
    $  3 ^ 3$  &      80.62    &     80.7559    &       -0.1359  \\
    $  5 ^ 3$  &     181.02    &    181.2481    &        -0.2281 \\
    $  7 ^ 3$  &     290.52    &    291.1644    &        -0.6444 \\
    $  9 ^ 3$  &     407.97    &    409.2399    &        -1.2699 \\
    $  4 ^ 4$  &     120.37    &    120.1674    &       0.2026   \\ 
    $  6 ^ 4$  &     228.23    &    228.4397    &       -0.2097  \\ 
    $  8 ^ 4$  &     345.27    &    344.4927    &       0.7773   \\
    \hline
  \end{tabular}
\end{minipage}
\hfill\vline\hfill
\begin{minipage}{0.4\textwidth}
  \begin{tabular}{cccc}
    \hline
    $n^{\ell}$ &        Exp. E. &      Calc. E. & Exp.-Calc. \\
    \hline 
    $ 10 ^ 4$  &     466.79    &    467.3681    &       -0.5781  \\
    $  5 ^ 5$  &     164.49    &    163.9965    &       0.4935   \\
    $  7 ^ 5$  &     278.61    &    278.8217    &        -0.2117 \\
    $  9 ^ 5$  &     401.59    &    400.1995    &       1.3905   \\
    $ 11 ^ 5$  &     528.08    &    527.3543    &       0.7257   \\
    $  6 ^ 6$  &     212.39    &    211.7692    &       0.6208   \\
    $  8 ^ 6$  &     331.89    &    332.2356    &       -0.3456  \\
    $ 10 ^ 6$  &     458.01    &    458.2881    &        -0.2781 \\
    $  7 ^ 7$  &     263.65    &    263.1749    &       0.4752   \\
    $  9 ^ 7$  &     388.95    &    388.5714    &       0.3786   \\
    $ 11 ^ 7$  &     518.15    &    518.7640    &       -0.6140  \\
    $  8 ^ 8$  &     317.94    &    318.0124    &       -0.0724  \\
    $ 10 ^ 8$  &     447.91    &    447.7643    &       0.1457   \\
    $  9 ^ 9$  &     375.65    &    376.1605    &       -0.5105  \\
    $ 11 ^ 9$  &     510.04    &    509.7902    &       0.2498   \\
    $ 10 ^{10}$  &     436.77    &    437.5598    &     -0.7898  \\
    $ 11 ^{11}$  &     501.91    &    502.2009    &      -0.2909 \\
    $ 12 ^{12}$  &     570.68    &    570.1168    &       0.5632 \\
    \hline
  \end{tabular}
\end{minipage}

    \caption{Experimental~\cite{OCCCO_2} and calculated term values and
      residuals for the CCC bending mode of OCCCO. Units of cm$^{-1}$.}
    \label{resOCCCO}
  \end{table}

\end{appendix}

% Provide your bibliography here. You have two options:

% FIRST OPTION - write your entries here directly, following the example below, including Author(s), Title, Journal Ref. with year in parentheses at the end, followed by the DOI number.
%\begin{thebibliography}{99}
%\bibitem{1931_Bethe_ZP_71} H. A. Bethe, {\it Zur Theorie der Metalle. i. Eigenwerte und Eigenfunktionen der linearen Atomkette}, Zeit. f{\"u}r Phys. {\bf 71}, 205 (1931), \doi{10.1007\%2FBF01341708}.
%\bibitem{arXiv:1108.2700} P. Ginsparg, {\it It was twenty years ago today... }, \url{http://arxiv.org/abs/1108.2700}.
%\end{thebibliography}

% \bibliographystyle{SciPost_bibstyle} % Include this style file here only if you are not using our template
\bibliography{mybibfile}

\begin{thebibliography}{100}
\providecommand{\url}[1]{\texttt{#1}}
\providecommand{\urlprefix}{URL }
\expandafter\ifx\csname urlstyle\endcsname\relax
  \providecommand{\doi}[1]{doi:\discretionary{}{}{}#1}\else
  \providecommand{\doi}{doi:\discretionary{}{}{}\begingroup
  \urlstyle{rm}\Url}\fi
\providecommand{\eprint}[2][]{\url{#2}}

\bibitem{Quapp1993}
W.~Quapp and B.~Winnewisser,
\newblock \emph{What you thought you already knew about the bending motion of
  triatomic molecules},
\newblock J.~Math.~Chem. \textbf{14}, 259 (1993),
\newblock \doi{https://doi.org/10.1007/BF01164471}.

\bibitem{Dixon1964}
R.~N. Dixon,
\newblock \emph{{H}igher {V}ibrational {L}evels of a {B}ent {T}riatomic
  {M}olecule},
\newblock Trans. Faraday Soc. \textbf{60}, 1363 (1964),
\newblock \doi{10.1039/TF9646001363}.

\bibitem{Thorson1960}
W.~Thorson and I.~Nakagawa,
\newblock \emph{{D}ynamics of the {Q}uasi‐{L}inear {M}olecule},
\newblock J. Chem. Phys. \textbf{33}, 994 (1960),
\newblock \doi{10.1063/1.1731399}.

\bibitem{Kroto1992}
H.~Kroto,
\newblock \emph{Molecular Rotation Spectra},
\newblock Dover (1992).

\bibitem{Winnewisser2006}
M.~Winnewisser, B.~Winnewisser, I.~Medvedev, F.~D. Lucia, S.~Ross and L.~Bates,
\newblock \emph{{The Hidden Kernel of Molecular Quasi-Linearity: Quantum
  Monodromy}},
\newblock J. Mol. Struct. \textbf{798}, 1  (2006),
\newblock \doi{https://doi.org/10.1016/j.molstruc.2006.06.036}.

\bibitem{Reilly2015}
N.~Reilly, P.~Changala, J.~Baraban, D.~Kokkin, J.~Stanton and M.~McCarthy,
\newblock \emph{Communication: The ground electronic state of {S}i$_2${C}:
  Rovibrational level structure, quantum monodromy, and astrophysical
  implications},
\newblock J.\ Chem.\ Phys. \textbf{142}, 231101 (2015),
\newblock \doi{10.1063/1.4922651}.

\bibitem{Zobov2005}
N.~Zobov, S.~Shirin, O.~Polyansky, J.~Tennyson, P.-F. Coheur, P.~Bernath,
  M.~Carleer and R.~Colin,
\newblock \emph{{Monodromy in the Water Molecule}},
\newblock Chem.\ Phys.\ Lett. \textbf{414}, 193  (2005),
\newblock \doi{https://doi.org/10.1016/j.cplett.2005.08.028}.

\bibitem{Winnewisser2005}
B.~Winnewisser, M.~Winnewisser, I.~Medvedev, M.~Behnke, F.~De~Lucia, S.~Ross
  and J.~Koput,
\newblock \emph{Experimental confirmation of quantum monodromy: The millimeter
  wave spectrum of cyanogen isothiocyanate {NCNCS}},
\newblock Phys. Rev. Lett. \textbf{95}, 243002 (2005),
\newblock \doi{10.1103/PhysRevLett.95.243002}.

\bibitem{Winnewisser2010}
B.~Winnewisser, M.~Winnewisser, I.~Medvedev, F.~De~Lucia, S.~Ross and J.~Koput,
\newblock \emph{{Analysis of the FASSST Rotational Spectrum of NCNCS in View of
  Quantum Monodromy}},
\newblock Phys. Chem. Chem. Phys. \textbf{12}, 8158 (2010),
\newblock \doi{10.1039/B922023B}.

\bibitem{Winnewisser2014}
M.~Winnewisser, B.~P. Winnewisser, F.~C. De~Lucia, D.~W. Tokaryk, S.~C. Ross
  and B.~E. Billinghurst,
\newblock \emph{{Pursuit of Quantum Monodromy in the Far-Infrared and
  Mid-Infrared Spectra of NCNCS Using Synchrotron Radiation}},
\newblock Phys. Chem. Chem. Phys. \textbf{16}, 17373 (2014),
\newblock \doi{10.1039/C4CP01443J}.

\bibitem{Child1998}
M.~S. Child,
\newblock \emph{Quantum states in a champagne bottle},
\newblock J.\ Phys.\ A: Math.\ and General \textbf{31}, 657 (1998),
\newblock \doi{10.1088/0305-4470/31/2/022}.

\bibitem{Child1999}
M.~S. Child, T.~Weston and J.~Tennyson,
\newblock \emph{Quantum monodromy in the spectrum of {H}$_2${O} and other
  systems: New insight into the level structure of quasi-linear molecules},
\newblock Mol.\ Phys. \textbf{96}, 371 (1999),
\newblock \doi{10.1080/00268979909482971}.

\bibitem{Duistermaat1980}
J.~J. Duistermaat,
\newblock \emph{On global action-angle coordinates},
\newblock Communications on Pure and Applied Mathematics \textbf{33}(6), 687
  (1980),
\newblock \doi{https://doi.org/10.1002/cpa.3160330602}.

\bibitem{HBJ}
J.~T. Hougen, P.~Bunker and J.~Johns,
\newblock \emph{The vibration-rotation problem in triatomic molecules allowing
  for a large-amplitude bending vibration},
\newblock J.\ Mol.\ Spectrosc. \textbf{34}, 136 (1970),
\newblock \doi{https://doi.org/10.1016/0022-2852(70)90080-9Q}.

\bibitem{Bunker1977}
P.~Bunker and B.~Landsberg,
\newblock \emph{The {R}igid {B}ender and {S}emirigid {B}ender {M}odels for the
  {R}otation-{V}ibration {H}amiltonian},
\newblock J. Mol. Spectrosc. \textbf{67}, 374  (1977),
\newblock \doi{https://doi.org/10.1016/0022-2852(77)90048-0}.

\bibitem{Ross1988}
S.~Ross,
\newblock \emph{{OCCCS}, {NCNCS}, {NCNCO}, and {NCNNN} as {S}emirigid
  {B}enders},
\newblock J.\ Mol.\ Spectrosc. \textbf{132}, 48  (1988),
\newblock \doi{https://doi.org/10.1016/0022-2852(88)90059-8}.

\bibitem{MORBID}
P.~Jensen,
\newblock \emph{A new morse oscillator-rigid bender internal dynamics
  ({MORBID}) {H}amiltonian for triatomic molecules},
\newblock J.\ Mol.\ Spectrosc. \textbf{128}, 478 (1988),
\newblock \doi{https://doi.org/10.1016/0022-2852(88)90164-6}.

\bibitem{Iachello2014}
F.~Iachello,
\newblock \emph{Lie Algebras and Applications},
\newblock Lecture Notes in Physics. Springer Berlin Heidelberg (2014).

\bibitem{general0}
F.~Iachello,
\newblock \emph{Lie algebras, cohomologies and new applications of quantum
  mechanics},
\newblock In N.~Kamran and P.~Olver, eds., \emph{Contemporary Mathematics},
  vol. 160, pp. 151--171. American Mathematical Society, Providence, RI (1994).

\bibitem{Bijker1994}
R.~Bijker, F.~Iachello and A.~Leviatan,
\newblock \emph{{Algebraic Models of Hadron Structure. I. Nonstrange Baryons
  }},
\newblock Ann.\ Phys. \textbf{236}, 69  (1994),
\newblock \doi{http://dx.doi.org/10.1006/aphy.1994.1108}.

\bibitem{Bijker2000}
R.~Bijker, F.~Iachello and A.~Leviatan,
\newblock \emph{Algebraic models of hadron structure: Ii. strange baryons},
\newblock Annals of Physics \textbf{284}(1), 89  (2000),
\newblock \doi{https://doi.org/10.1006/aphy.2000.6064}.

\bibitem{booknuc}
F.~Iachello and A.~Arima,
\newblock \emph{The Interacting Boson Model},
\newblock Cambridge University Press, Cambridge,
\newblock \doi{https://doi.org/10.1017/CBO9780511895517} (1987).

\bibitem{Iachello1991book}
F.~Iachello, P.~van Isacker, P.~Van~Isacker, P.~Landshoff, D.~Nelson, D.~Sciama
  and S.~Weinberg,
\newblock \emph{The Interacting Boson-Fermion Model},
\newblock Cambridge University Press (1991).

\bibitem{frank}
A.~Frank and P.~V. Isacker,
\newblock \emph{Algebraic Methods in Molecular and Nuclear Structure Physics},
\newblock John Wiley and Sons, New York (1994).

\bibitem{Iachello:81}
F.~Iachello,
\newblock \emph{Algebraic methods for molecular rotation-vibration spectra},
\newblock Chem.\ Phys.\ Lett. \textbf{78}, 581  (1981),
\newblock \doi{http://dx.doi.org/10.1016/0009-2614(81)85262-1}.

\bibitem{bookmol}
F.~Iachello and R.~Levine,
\newblock \emph{Algebraic Theory of Molecules},
\newblock Oxford University Press, Oxford,
\newblock \doi{https://doi.org/10.1002/bbpc.19950990427} (1995).

\bibitem{Iachello1996}
F.~Iachello and S.~Oss,
\newblock \emph{Algebraic approach to molecular spectra: Two dimensional
  problems},
\newblock J.~Chem.~Phys. \textbf{104}, 6956 (1996),
\newblock \doi{http://dx.doi.org/10.1063/1.471412}.

\bibitem{PBernal2008}
F.~P\'erez-Bernal and F.~Iachello,
\newblock \emph{Algebraic approach to two-dimensional systems: Shape phase
  transitions, monodromy, and thermodynamic quantities},
\newblock Phys. Rev. A \textbf{77}, 032 (2008),
\newblock \doi{10.1103/PhysRevA.77.032115}.

\bibitem{Iachello2003}
F.~Iachello, F.~P\'erez-Bernal and P.~Vaccaro,
\newblock \emph{{A Novel Algebraic Scheme for Describing Nonrigid Molecules}},
\newblock Chem.\ Phys.\ Lett. \textbf{375}, 309  (2003),
\newblock \doi{https://doi.org/10.1016/S0009-2614(03)00851-0}.

\bibitem{PBernal2005}
F.~P\'erez-Bernal, L.~F. Santos, P.~Vaccaro and F.~Iachello,
\newblock \emph{{Spectroscopic Signatures of Nonrigidity: Algebraic Analyses of
  Infrared and {R}aman Transitions in Nonrigid Species}},
\newblock Chem.\ Phys.\ Lett. \textbf{414}, 398  (2005),
\newblock \doi{https://doi.org/10.1016/j.cplett.2005.07.119}.

\bibitem{Larese2011}
D.~Larese and F.~Iachello,
\newblock \emph{{A Study of Quantum Phase Transitions and Quantum Monodromy in
  the Bending Motion of Non-Rigid Molecules}},
\newblock J. Mol. Struct. \textbf{1006}, 611  (2011),
\newblock \doi{https://doi.org/10.1016/j.molstruc.2011.10.016}.

\bibitem{Larese2013}
D.~Larese, F.~P\'erez-Bernal and F.~Iachello,
\newblock \emph{{Signatures of Quantum Phase Transitions and Excited State
  Quantum Phase Transitions in the Vibrational Bending Dynamics of Triatomic
  Molecules}},
\newblock J. Mol. Struct. \textbf{1051}, 310  (2013),
\newblock \doi{https://doi.org/10.1016/j.molstruc.2013.08.020}.

\bibitem{KRivera2020}
J.~Khalouf-Rivera, F.~Pérez-Bernal and M.~Carvajal,
\newblock \emph{Excited state quantum phase transitions in the bending spectra
  of molecules},
\newblock J. Quant. Spectrosc. and Rad. Transfer p. 107436 (2020),
\newblock \doi{https://doi.org/10.1016/j.jqsrt.2020.107436}.

\bibitem{Iachello2008}
F.~Iachello and F.~Pérez-bernal,
\newblock \emph{Bending vibrational modes of {ABBA} molecules: Algebraic
  approach and its classical limit},
\newblock Mol. Phys. \textbf{106}(2-4), 223 (2008),
\newblock \doi{10.1080/00268970701757883}.

\bibitem{Iachello2009}
F.~Iachello and F.~P\'erez-Bernal,
\newblock \emph{A novel algebraic scheme for describing coupled benders in
  tetratomic molecules},
\newblock J. Phys. Chem. A \textbf{113}, 13273 (2009),
\newblock \doi{10.1021/jp9040474}.

\bibitem{Larese2014}
D.~Larese, M.~Caprio, F.~P\'erez-Bernal and F.~Iachello,
\newblock \emph{A study of the bending motion in tetratomic molecules by the
  algebraic operator expansion method},
\newblock J. Chem. Phys. \textbf{140}, 014 (2014),
\newblock \doi{10.1063/1.4856115}.

\bibitem{Ishikawa2002}
H.~Ishikawa, H.~Toyosaki, N.~Mikami, F.~P\'erez-Bernal, P.~Vaccaro and
  F.~Iachello,
\newblock \emph{{Algebraic Analysis of Bent-from-Linear Transition Intensities:
  the Vibronically Resolved Emission Spectrum of Methinophosphide ({HCP})}},
\newblock Chem.\ Phys.\ Lett. \textbf{365}, 57  (2002),
\newblock \doi{https://doi.org/10.1016/S0009-2614(02)01419-7}.

\bibitem{SCastellanos2012}
M.~S\'anchez-Castellanos, R.~Lemus, M.~Carvajal, F.~P\'erez-Bernal and J.~M.
  Fern\'andez,
\newblock \emph{{A Study of the Raman Spectrum of {CO}$_2$ Using an Algebraic
  Approach}},
\newblock Chem.\ Phys.\ Lett. \textbf{554}, 208  (2012),
\newblock \doi{https://doi.org/10.1016/j.cplett.2012.09.075}.

\bibitem{Lemus2014}
R.~Lemus, M.~S\'anchez-Castellanos, F.~P\'erez-Bernal, J.~M. Fern\'andez and
  M.~Carvajal,
\newblock \emph{Simulation of the {R}aman spectra of {CO}$_2$: Bridging the gap
  between algebraic models and experimental spectra},
\newblock J.\ Chem.\ Phys. \textbf{141}, 054 (2014),
\newblock \doi{10.1063/1.4889995}.

\bibitem{Marisol2020}
M.~Berm\'udez-Monta{\~n}a, M.~Carvajal, F.~P\'erez-Bernal and R.~Lemus,
\newblock \emph{An algebraic alternative for the accurate simulation of
  {CO}$_2$ {R}aman spectra},
\newblock J.\ Raman Spectrosc. \textbf{51}, 569 (2020),
\newblock \doi{10.1002/jrs.5801}.

\bibitem{KRivera2019}
J.~Khalouf-Rivera, M.~Carvajal, L.~F. Santos and F.~P\'erez-Bernal,
\newblock \emph{Calculation of transition state energies in the {HCN–HNC}
  isomerization with an algebraic model},
\newblock J.\ Phys.\ Chem.\ A \textbf{123}, 9544 (2019),
\newblock \doi{10.1021/acs.jpca.9b07338}.

\bibitem{Carr2010}
L.~Carr,
\newblock \emph{Understanding Quantum Phase Transitions},
\newblock Condensed Matter Physics. CRC Press,
\newblock ISBN 9781439802618 (2010).

\bibitem{Sachdev2011}
S.~Sachdev,
\newblock \emph{Quantum Phase Transitions},
\newblock Cambridge University Press, Cambridge,
\newblock \doi{https://doi.org/10.1017/CBO9780511973765} (2011).

\bibitem{Gilmore1978}
R.~Gilmore and D.~Feng,
\newblock \emph{Phase transitions in nuclear matter described by pseudospin
  {H}amiltonians},
\newblock Nucl. Phys. A \textbf{301}(2), 189  (1978),
\newblock \doi{https://doi.org/10.1016/0375-9474(78)90260-9}.

\bibitem{Gilmore1979}
R.~Gilmore,
\newblock \emph{The classical limit of quantum nonspin systems},
\newblock J.\ Math.\ Phys. \textbf{20}, 891 (1979),
\newblock \doi{10.1063/1.524137}.

\bibitem{Feng1981}
D.~Feng, R.~Gilmore and S.~Deans,
\newblock \emph{Phase transitions and the geometric properties of the
  interacting boson model},
\newblock Phys. Rev. C \textbf{23}, 1254 (1981),
\newblock \doi{10.1103/PhysRevC.23.1254}.

\bibitem{Iachello2004}
F.~Iachello and N.~V. Zamfir,
\newblock \emph{Quantum phase transitions in mesoscopic systems},
\newblock Phys. Rev. Lett. \textbf{92}, 212501 (2004),
\newblock \doi{10.1103/PhysRevLett.92.212501}.

\bibitem{Casten2009}
R.~Casten,
\newblock \emph{{Quantum Phase Transitions and Structural Evolution in
  Nuclei}},
\newblock Prog.\ Part.\ Nucl.\ Phys. \textbf{62}, 183  (2009),
\newblock \doi{https://doi.org/10.1016/j.ppnp.2008.06.002}.

\bibitem{Cejnar2009}
P.~Cejnar and J.~Jolie,
\newblock \emph{{Quantum Phase Transitions in the Interacting Boson Model}},
\newblock Prog. Part. Nucl. Phys. \textbf{62}, 210  (2009),
\newblock \doi{https://doi.org/10.1016/j.ppnp.2008.08.001}.

\bibitem{Cejnar2010}
P.~Cejnar, J.~Jolie and R.~Casten,
\newblock \emph{Quantum phase transitions in the shapes of atomic nuclei},
\newblock Rev. Mod. Phys. \textbf{82}, 2155 (2010),
\newblock \doi{10.1103/RevModPhys.82.2155}.

\bibitem{Cejnar2007}
P.~Cejnar and F.~Iachello,
\newblock \emph{Phase structure of interacting boson models in arbitrary
  dimension},
\newblock J. Phys. A: Math. and Theor. \textbf{40}, 581 (2007),
\newblock \doi{https://doi.org/10.1088/1751-8113/40/4/001}.

\bibitem{PFernandez2011}
P.~P\'erez-Fern\'andez, J.~Arias, J.~E. Garc\'{\i}a-Ramos and
  F.~P\'erez-Bernal,
\newblock \emph{Finite-size corrections in the bosonic algebraic approach to
  two-dimensional systems},
\newblock Phys. Rev. A \textbf{83}, 062125 (2011),
\newblock \doi{10.1103/PhysRevA.83.062125}.

\bibitem{Zhang2010}
Y.~Zhang, F.~Pan, Y.-X. Liu and J.~Draayer,
\newblock \emph{The {E}(2) symmetry and quantum phase transition in the
  two-dimensional limit of the vibron model},
\newblock J. Phys. B -- At. Mol. Opt. \textbf{43}, 225101 (2010),
\newblock \doi{10.1088/0953-4075/43/22/225101}.

\bibitem{Calixto2012}
M.~Calixto, E.~Romera and R.~del Real,
\newblock \emph{Parity-symmetry-adapted coherent states and entanglement in
  quantum phase transitions of vibron models},
\newblock J. Phys. A: Math. Theor. \textbf{45}, 365301 (2012),
\newblock \doi{10.1088/1751-8113/45/36/365301}.

\bibitem{Calixto2012b}
M.~Calixto, R.~del Real and E.~Romera,
\newblock \emph{Husimi distribution and phase-space analysis of a vibron-model
  quantum phase transition},
\newblock Phys. Rev. A \textbf{86}, 032508 (2012),
\newblock \doi{10.1103/PhysRevA.86.032508}.

\bibitem{Santos2013}
F.~de~los Santos and E.~Romera,
\newblock \emph{Revival times at quantum phase transitions},
\newblock Phys. Rev. A \textbf{87}, 013424 (2013),
\newblock \doi{10.1103/PhysRevA.87.013424}.

\bibitem{Castanos2015}
O.~Casta\~nos, M.~Calixto, F.~P\'erez-Bernal and E.~Romera,
\newblock \emph{Identifying the order of a quantum phase transition by means of
  wehrl entropy in phase space},
\newblock Phys. Rev. E \textbf{92}, 052106 (2015),
\newblock \doi{10.1103/PhysRevE.92.052106}.

\bibitem{Cejnar2006}
P.~Cejnar, M.~Macek, S.~Heinze, J.~Jolie and J.~Dobe\v{s},
\newblock \emph{Monodromy and excited-state quantum phase transitions in
  integrable systems: Collective vibrations of nuclei},
\newblock J.\ Phys. A: Math.\ and General \textbf{39}, L515 (2006),
\newblock \doi{10.1088/0305-4470/39/31/l01}.

\bibitem{Cejnar2007b}
P.~Cejnar, S.~Heinze and M.~Macek,
\newblock \emph{Coulomb analogy for non-hermitian degeneracies near quantum
  phase transitions},
\newblock Phys. Rev. Lett. \textbf{99}, 100601 (2007),
\newblock \doi{10.1103/PhysRevLett.99.100601}.

\bibitem{Caprio2008}
M.~Caprio, P.~Cejnar and F.~Iachello,
\newblock \emph{{Excited State Quantum Phase Transitions in Many-Body
  Systems}},
\newblock Ann.\ Phys. \textbf{323}, 1106  (2008),
\newblock \doi{https://doi.org/10.1016/j.aop.2007.06.011}.

\bibitem{Fernandez2009}
P.~P\'erez-Fern\'andez, A.~Rela\~no, J.~M. Arias, J.~Dukelsky and J.~E.
  Garc\'{i}a-Ramos,
\newblock \emph{Decoherence due to an excited-state quantum phase transition in
  a two-level boson model},
\newblock Phys. Rev. A \textbf{80}, 032111 (2009),
\newblock \doi{10.1103/PhysRevA.80.032111}.

\bibitem{Fernandez2011b}
P.~P\'erez-Fern\'andez, A.~Rela\~no, J.~M. Arias, P.~Cejnar, J.~Dukelsky and
  J.~E. Garc\'{i}a-Ramos,
\newblock \emph{Excited-state phase transition and onset of chaos in quantum
  optical models},
\newblock Phys. Rev. E \textbf{83}, 046208 (2011),
\newblock \doi{10.1103/PhysRevE.83.046208}.

\bibitem{Cejnar2008}
P.~Cejnar and P.~Stransky,
\newblock \emph{Impact of quantum phase transitions on excited-level dynamics},
\newblock Phys.\ Rev.\ E \textbf{78} (2008),
\newblock \doi{10.1103/PhysRevE.78.031130}.

\bibitem{Stransky2014}
P.~Stránský, M.~Macek and P.~Cejnar,
\newblock \emph{{Excited-State Quantum Phase Transitions in Systems with Two
  Degrees of Freedom: Level Density, Level Dynamics, Thermal Properties}},
\newblock Ann. Phys. \textbf{345}, 73  (2014),
\newblock \doi{https://doi.org/10.1016/j.aop.2014.03.006}.

\bibitem{Stransky2015}
P.~Stránský, M.~Macek, A.~Leviatan and P.~Cejnar,
\newblock \emph{{Excited-State Quantum Phase Transitions in Systems with Two
  Degrees of Freedom: II. Finite-Size Effects}},
\newblock Ann. Phys. \textbf{356}, 57  (2015),
\newblock \doi{https://doi.org/10.1016/j.aop.2015.02.025}.

\bibitem{Macek2019}
M.~Macek, P.~Str\'ansk\'y, A.~Leviatan and P.~Cejnar,
\newblock \emph{Excited-state quantum phase transitions in systems with two
  degrees of freedom. {III}. interacting boson systems},
\newblock Phys. Rev. C \textbf{99}, 064323 (2019),
\newblock \doi{10.1103/PhysRevC.99.064323}.

\bibitem{Cejnar2020}
P.~Cejnar, P.~Stránský, M.~Macek and M.~Kloc,
\newblock \emph{Excited-state quantum phase transitions},
\newblock J. Phys. A: Mathem. and Theoret.  (2021).

\bibitem{GRamos2017}
J.~E. Garc\'{\i}a-Ramos, P.~P\'erez-Fern\'andez and J.~M. Arias,
\newblock \emph{Excited-state quantum phase transitions in a two-fluid {L}ipkin
  model},
\newblock Phys. Rev. C \textbf{95}, 054326 (2017),
\newblock \doi{10.1103/PhysRevC.95.054326}.

\bibitem{Relano2016}
A.~Relaño, C.~Esebbag and J.~Dukelsky,
\newblock \emph{Excited-state quantum phase transitions in the two-spin
  elliptic gaudin model},
\newblock Physical Review E \textbf{94}, 052110 (2016),
\newblock \doi{10.1103/PhysRevE.94.052110}.

\bibitem{Brandes2013}
T.~Brandes,
\newblock \emph{Excited-state quantum phase transitions in {D}icke
  superradiance models},
\newblock Phys.\ Rev.\ E \textbf{88}, 032133 (2013),
\newblock \doi{10.1103/PhysRevE.88.032133}.

\bibitem{Bastidas2014}
V.~Bastidas, P.~P\'erez-Fern\'andez, M.~Vogl and T.~Brandes,
\newblock \emph{Quantum criticality and dynamical instability in the kicked-top
  model},
\newblock Phys.\ Rev.\ Lett. \textbf{112}, 140408 (2014),
\newblock \doi{10.1103/PhysRevLett.112.140408}.

\bibitem{Dietz2013}
B.~Dietz, F.~Iachello, M.~Miski-Oglu, N.~Pietralla, A.~Richter, L.~von Smekal
  and J.~Wambach,
\newblock \emph{Lifshitz and excited-state quantum phase transitions in
  microwave dirac billiards},
\newblock Phys. Rev. B \textbf{88}, 104101 (2013),
\newblock \doi{10.1103/PhysRevB.88.104101}.

\bibitem{Dietz2017}
B.~Dietz, F.~Iachello and M.~Macek,
\newblock \emph{Algebraic theory of crystal vibrations: Localization properties
  of wave functions in two-dimensional lattices},
\newblock Crystals \textbf{7}, 246 (2017),
\newblock \doi{10.3390/cryst7080246}.

\bibitem{Feldmann2020}
P.~Feldmann, C.~Klempt, A.~Smerzi, L.~Santos and M.~Gessner,
\newblock \emph{Excited-state quantum phase transitions in spinor
  {B}ose-{E}instein condensates} (2020), \eprint{arXiv:2011.02823}.

\bibitem{Cabedo2021}
J.~Cabedo, J.~Claramunt and A.~Celi,
\newblock \emph{Excited-state quantum phase transitions in spin-orbit coupled
  bose gases} (2021), \eprint{2101.08253}.

\bibitem{Puebla2013}
R.~Puebla, A.~Rela\~no and J.~Retamosa,
\newblock \emph{Excited-state phase transition leading to symmetry-breaking
  steady states in the {D}icke model},
\newblock Phys. Rev. A \textbf{87}, 023819 (2013),
\newblock \doi{10.1103/PhysRevA.87.023819}.

\bibitem{Puebla2015}
R.~Puebla and A.~Rela\~no,
\newblock \emph{Irreversible processes without energy dissipation in an
  isolated {Lipkin-Meshkov-Glick} model},
\newblock Phys. Rev. E \textbf{92}, 012101 (2015),
\newblock \doi{10.1103/PhysRevE.92.012101}.

\bibitem{Bastidas2015}
G.~Engelhardt, V.~M. Bastidas, W.~Kopylov and T.~Brandes,
\newblock \emph{Excited-state quantum phase transitions and periodic dynamics},
\newblock Phys. Rev. A \textbf{91}, 013631 (2015),
\newblock \doi{10.1103/PhysRevA.91.013631}.

\bibitem{Santos2015}
L.~F. Santos and F.~P\'erez-Bernal,
\newblock \emph{Structure of eigenstates and quench dynamics at an
  excited-state quantum phase transition},
\newblock Phys. Rev. A \textbf{92}, 050101 (2015),
\newblock \doi{10.1103/PhysRevA.92.050101}.

\bibitem{Santos2016}
L.~F. Santos, M.~T\'avora and F.~P\'erez-Bernal,
\newblock \emph{Excited-state quantum phase transitions in many-body systems
  with infinite-range interaction: Localization, dynamics, and bifurcation},
\newblock Phys. Rev. A \textbf{94}, 012 (2016),
\newblock \doi{10.1103/PhysRevA.94.012113}.

\bibitem{PBernal2017}
F.~P\'erez-Bernal and L.~F. Santos,
\newblock \emph{Effects of excited state quantum phase transitions on system
  dynamics},
\newblock Progr. Phys. Fortschr. Phys. \textbf{65}(6-8), 1600035 (2017),
\newblock \doi{10.1002/prop.201600035}.

\bibitem{Wang2017}
Q.~Wang and H.~T. Quan,
\newblock \emph{Probing the excited-state quantum phase transition through
  statistics of loschmidt echo and quantum work},
\newblock Phys. Rev. E \textbf{96}, 032142 (2017),
\newblock \doi{10.1103/PhysRevE.96.032142}.

\bibitem{Kopylov2017}
W.~Kopylov, G.~Schaller and T.~Brandes,
\newblock \emph{Nonadiabatic dynamics of the excited states for the
  {Lipkin-Meshkov-Glick} model},
\newblock Phys. Rev. E \textbf{96}, 012153 (2017),
\newblock \doi{10.1103/PhysRevE.96.012153}.

\bibitem{Kloc2018}
M.~Kloc, P.~Str\'ansk\'y and P.~Cejnar,
\newblock \emph{Quantum quench dynamics in {D}icke superradiance models},
\newblock Phys. Rev. A \textbf{98}, 013836 (2018),
\newblock \doi{10.1103/PhysRevA.98.013836}.

\bibitem{Wang2019a}
Q.~Wang and F.~P\'erez-Bernal,
\newblock \emph{Excited-state quantum phase transition and the
  quantum-speed-limit time},
\newblock Phys. Rev. A \textbf{100}, 022118 (2019),
\newblock \doi{10.1103/PhysRevA.100.022118}.

\bibitem{Wang2019b}
Q.~Wang and F.~P\'erez-Bernal,
\newblock \emph{Probing an excited-state quantum phase transition in a quantum
  many-body system via an out-of-time-order correlator},
\newblock Phys. Rev. A \textbf{100}, 062113 (2019),
\newblock \doi{10.1103/PhysRevA.100.062113}.

\bibitem{BMagnani2016}
M.~A. Bastarrachea-Magnani, S.~Lerma-Hern\'{a}ndez and J.~G. Hirsch,
\newblock \emph{Thermal and quantum phase transitions in atom-field systems: a
  microcanonical analysis},
\newblock J. Stat. Mech. Theory Exp. \textbf{2016}(9), 093105 (2016),
\newblock \doi{10.1088/1742-5468/2016/09/093105}.

\bibitem{PFernandez2017}
P.~P\'erez-Fern\'andez and A.~Rela\~no,
\newblock \emph{From thermal to excited-state quantum phase transition: The
  {D}icke model},
\newblock Phys. Rev. E \textbf{96}, 012121 (2017),
\newblock \doi{10.1103/PhysRevE.96.012121}.

\bibitem{Heinze2006}
S.~Heinze, P.~Cejnar, J.~Jolie and M.~Macek,
\newblock \emph{Evolution of spectral properties along the {O(6)-U(5)}
  transition in the interacting boson model. i. level dynamics},
\newblock Phys. Rev. C \textbf{73}, 014306 (2006),
\newblock \doi{10.1103/PhysRevC.73.014306}.

\bibitem{Macek2006}
M.~Macek, P.~Cejnar, J.~Jolie and S.~Heinze,
\newblock \emph{Evolution of spectral properties along the {O(6)-U(5)}
  transition in the interacting boson model. ii. classical trajectories},
\newblock Phys. Rev. C \textbf{73}, 014307 (2006),
\newblock \doi{10.1103/PhysRevC.73.014307}.

\bibitem{Kloc2017}
M.~Kloc, P.~Str\'{a}nsk\'{y} and P.~Cejnar,
\newblock \emph{Monodromy in {D}icke superradiance},
\newblock J.\ Phys.\ A: Math. and Theor. \textbf{50}(31), 315205 (2017),
\newblock \doi{10.1088/1751-8121/aa7a95}.

\bibitem{Zhao2014}
L.~Zhao, J.~Jiang, T.~Tang, M.~Webb and Y.~Liu,
\newblock \emph{Dynamics in spinor condensates tuned by a microwave dressing
  field},
\newblock Phys. Rev. A \textbf{89}, 023608 (2014),
\newblock \doi{10.1103/PhysRevA.89.023608}.

\bibitem{Evers2008}
F.~Evers and A.~Mirlin,
\newblock \emph{Anderson transitions},
\newblock Rev. Mod. Phys. \textbf{80}, 1355 (2008),
\newblock \doi{10.1103/RevModPhys.80.1355}.

\bibitem{Izrailev1990}
F.~Izrailev,
\newblock \emph{{Simple Models of Quantum Chaos: Spectrum and Eigenfunctions}},
\newblock Phys. Rep. \textbf{196}, 299  (1990),
\newblock \doi{https://doi.org/10.1016/0370-1573(90)90067-C}.

\bibitem{Zelevinsky1996}
V.~Zelevinsky, B.~Brown, N.~Frazier and M.~Horoi,
\newblock \emph{{The Nuclear Shell Model as a Testing Ground for Many-Body
  Quantum Chaos}},
\newblock Phys.\ Rep. \textbf{276}, 85  (1996),
\newblock \doi{https://doi.org/10.1016/S0370-1573(96)00007-5}.

\bibitem{Rowe2004}
D.~J. Rowe,
\newblock \emph{Quasidynamical symmetry in an interacting boson model phase
  transition},
\newblock Phys. Rev. Lett. \textbf{93}, 122502 (2004),
\newblock \doi{10.1103/PhysRevLett.93.122502}.

\bibitem{Amico2008}
L.~Amico, R.~Fazio, A.~Osterloh and V.~Vedral,
\newblock \emph{Entanglement in many-body systems},
\newblock Rev. Mod. Phys. \textbf{80}, 517 (2008),
\newblock \doi{10.1103/RevModPhys.80.517}.

\bibitem{Gu2010}
S.-J. Gu,
\newblock \emph{Fidelity approach to quantum phase transitions},
\newblock International Journal of Modern Physics B \textbf{24}(23), 4371
  (2010),
\newblock \doi{10.1142/S0217979210056335}.

\bibitem{Braun2018}
D.~Braun, G.~Adesso, F.~Benatti, R.~Floreanini, U.~Marzolino, M.~W. Mitchell
  and S.~Pirandola,
\newblock \emph{Quantum-enhanced measurements without entanglement},
\newblock Rev. Mod. Phys. \textbf{90}, 035006 (2018),
\newblock \doi{10.1103/RevModPhys.90.035006}.

\bibitem{Nielsen2000}
M.~A. Nielsen and I.~L. Chuang,
\newblock \emph{Quantum Computation and Quantum Information: 10th Anniversary
  Edition},
\newblock Cambridge University Press, USA, 10th edn.,
\newblock \doi{https://doi.org/10.1017/CBO9780511976667} (2011).

\bibitem{Zanardi2006}
P.~Zanardi and N.~Paunkovi\ifmmode~\acute{c}\else \'{c}\fi{},
\newblock \emph{Ground state overlap and quantum phase transitions},
\newblock Phys. Rev. E \textbf{74}, 031123 (2006),
\newblock \doi{10.1103/PhysRevE.74.031123}.

\bibitem{You2007}
W.-L. You, Y.-W. Li and S.-J. Gu,
\newblock \emph{Fidelity, dynamic structure factor, and susceptibility in
  critical phenomena},
\newblock Phys. Rev. E \textbf{76}, 022101 (2007),
\newblock \doi{10.1103/PhysRevE.76.022101}.

\bibitem{Gu2008}
S.-J. Gu, H.-M. Kwok, W.-Q. Ning and H.-Q. Lin,
\newblock \emph{Fidelity susceptibility, scaling, and universality in quantum
  critical phenomena},
\newblock Phys. Rev. B \textbf{77}, 245109 (2008),
\newblock \doi{10.1103/PhysRevB.77.245109}.

\bibitem{Santopinto1996}
E.~Santopinto, R.~Bijker and F.~Iachello,
\newblock \emph{{Transformation brackets between \(U(\nu+1)\supset U(\nu)
  \supset SO(\nu)\) and \(U(\nu+1)\supset SO(\nu+1)\supset SO(\nu)\)}},
\newblock J. of Math. Phys. \textbf{37}(6), 2674 (1996),
\newblock \doi{10.1063/1.531689}.

\bibitem{Carrasquilla2013}
J.~Carrasquilla, S.~R. Manmana and M.~Rigol,
\newblock \emph{Scaling of the gap, fidelity susceptibility, and bloch
  oscillations across the superfluid-to-mott-insulator transition in the
  one-dimensional {B}ose-{H}ubbard model},
\newblock Phys. Rev. A \textbf{87}, 043606 (2013),
\newblock \doi{10.1103/PhysRevA.87.043606}.

\bibitem{Yang2008}
S.~Yang, S.-J. {G}u, C.-P. Sun and H.-Q. Lin,
\newblock \emph{Fidelity susceptibility and long-range correlation in the
  {K}itaev honeycomb model},
\newblock Phys. Rev. A \textbf{78}, 012304 (2008),
\newblock \doi{10.1103/PhysRevA.78.012304}.

\bibitem{Kwok2008}
H.-M. Kwok, W.-Q. Ning, S.-J. Gu and H.-Q. Lin,
\newblock \emph{Quantum criticality of the {Lipkin-Meshkov-Glick} model in
  terms of fidelity susceptibility},
\newblock Phys. Rev. E \textbf{78}, 032103 (2008),
\newblock \doi{10.1103/PhysRevE.78.032103}.

\bibitem{Leung2012}
C.-Y. Leung, W.~C. {Y}u, H.-M. Kwok, S.-J. {G}u and H.-Q. Lin,
\newblock \emph{Scaling behavior of the ground-state fidelity in the
  {L}ipkin-{M}eshkov-{G}lick model},
\newblock Int. J. Modern Phys. B \textbf{26}(31), 1250170 (2012),
\newblock \doi{10.1142/S0217979212501706}.

\bibitem{Romera2014}
E.~Romera, M.~Calixto and O.~C. nos,
\newblock \emph{Phase space analysis of first-, second- and third-order quantum
  phase transitions in the {Lipkin–Meshkov–Glick} model},
\newblock Physica Scripta \textbf{89}, 095103 (2014).

\bibitem{Yu2009}
W.-C. Yu, H.-M. Kwok, J.~Cao and S.-J. Gu,
\newblock \emph{Fidelity susceptibility in the two-dimensional transverse-field
  ising and $xxz$ models},
\newblock Phys. Rev. E \textbf{80}, 021108 (2009),
\newblock \doi{10.1103/PhysRevE.80.021108}.

\bibitem{Wei2018}
B.-B. Wei and X.-C. Lv,
\newblock \emph{Fidelity susceptibility in the quantum {R}abi model},
\newblock Phys. Rev. A \textbf{97}, 013845 (2018),
\newblock \doi{10.1103/PhysRevA.97.013845}.

\bibitem{Sierant2019}
P.~Sierant, A.~Maksymov, M.~Ku\ifmmode~\acute{s}\else \'{s}\fi{} and
  J.~Zakrzewski,
\newblock \emph{Fidelity susceptibility in gaussian random ensembles},
\newblock Phys. Rev. E \textbf{99}, 050102 (2019),
\newblock \doi{10.1103/PhysRevE.99.050102}.

\bibitem{Rams2011}
M.~M. Rams and B.~Damski,
\newblock \emph{Quantum fidelity in the thermodynamic limit},
\newblock Phys. Rev. Lett. \textbf{106}, 055701 (2011),
\newblock \doi{10.1103/PhysRevLett.106.055701}.

\bibitem{Greschner2013}
S.~Greschner, A.~K. Kolezhuk and T.~Vekua,
\newblock \emph{Fidelity susceptibility and conductivity of the current in
  one-dimensional lattice models with open or periodic boundary conditions},
\newblock Phys. Rev. B \textbf{88}, 195101 (2013),
\newblock \doi{10.1103/PhysRevB.88.195101}.

\bibitem{Wei2020}
B.-B. Wei,
\newblock \emph{Fidelity susceptibility in one-dimensional disordered lattice
  models},
\newblock Phys. Rev. A \textbf{99}, 042117 (2019),
\newblock \doi{10.1103/PhysRevA.99.042117}.

\bibitem{Leblond2020}
T.~LeBlond, D.~Sels, A.~Polkovnikov and M.~Rigol,
\newblock \emph{Universality in the onset of quantum chaos in many-body
  systems} (2020), \eprint{arXiv: 2012.07849}.

\bibitem{PBernal2010}
F.~P\'erez-Bernal and O.~\'Alvarez-Bajo,
\newblock \emph{Anharmonicity effects in the bosonic {U(2)-SO(3)} excited-state
  quantum phase transition},
\newblock Phys. Rev. A \textbf{81}, 050 (2010),
\newblock \doi{10.1103/PhysRevA.81.050101}.

\bibitem{MellauHNC}
G.~Mellau,
\newblock \emph{Complete experimental rovibrational eigenenergies of {HNC} up
  to 3743cm$^{-1}$ above the ground state},
\newblock J.\ Chem.\ Phys. \textbf{133}, 164 (2010),
\newblock \doi{10.1063/1.3503508}.

\bibitem{Koput1986}
J.~Koput,
\newblock \emph{The microwave-spectrum of methyl isocyanate},
\newblock J.\ Mol.\ Spectrosc. \textbf{115}, 131 (1986),
\newblock \doi{10.1016/0022-2852(86)90281-X}.

\bibitem{Br_Cl_CNO2001}
H.~Lichau, C.~Gillies, J.~Gillies, S.~Ross, B.~Winnewisser and M.~Winnewisser,
\newblock \emph{{On the anharmonic XCN bending modes of the quasilinear
  molecules BrCNO and ClCNO}},
\newblock J.\ Phys.\ Chem.\ A \textbf{105}, 10065 (2001),
\newblock \doi{10.1021/jp012067u}.

\bibitem{OCCCO_2}
J.~VanderAuwera, J.~Johns and O.~Polyansky,
\newblock \emph{The far infrared-spectrum of {C$_3$O$_2$}},
\newblock J.\ Chem.\ Phys. \textbf{95}(4), 2299 (1991).

\end{thebibliography}

\nolinenumbers

\end{document}